\documentclass[
        pra,
	aps,
	amsmath,amssymb,
	reprint,
	superscriptaddress,
]{revtex4-2}

\usepackage[dvipdfmx]{graphicx}
\usepackage{color}
\usepackage{dcolumn}
\usepackage{siunitx}
\usepackage{bm}
\usepackage{braket}
\usepackage{tikz}
\usepackage{qcircuit}
\usepackage{enumitem}
\usepackage{booktabs, tabularx}
\usepackage{mathtools}
\usepackage[normalem]{ulem}
\usepackage[breaklinks]{hyperref}

\newcommand{\cmidruleleft}{\cmidrule(lr){2-4}}
\newcommand{\cmidruleright}{\cmidrule(lr){5-7}}

\begin{document}

\title[]{Measurement-Based Fault-Tolerant Quantum Computation on High-Connectivity Devices: A Resource-Efficient Approach toward Early FTQC}

\author{Yohei Ibe}
\email{ibe@qunasys.com}
\affiliation{QunaSys Inc., Aqua Hakusan Building 9F, 1-13-7 Hakusan, Bunkyo, Tokyo 113-0001, Japan}

\author{Yutaka Hirano}
\affiliation{Graduate School of Engineering Science, The University of Osaka, 1-3 Machikaneyama, Toyonaka, Osaka 560-8531, Japan}

\author{Yasuo Ozu}
\affiliation{QunaSys Inc., Aqua Hakusan Building 9F, 1-13-7 Hakusan, Bunkyo, Tokyo 113-0001, Japan}

\author{Toru Kawakubo}
\affiliation{QunaSys Inc., Aqua Hakusan Building 9F, 1-13-7 Hakusan, Bunkyo, Tokyo 113-0001, Japan}

\author{Keisuke Fujii}
\affiliation{QunaSys Inc., Aqua Hakusan Building 9F, 1-13-7 Hakusan, Bunkyo, Tokyo 113-0001, Japan}
\affiliation{Graduate School of Informatics, Kyoto University, Sakyo-ku, Kyoto, 606-8501, Japan}
\affiliation{Graduate School of Engineering Science, The University of Osaka, 1-3 Machikaneyama, Toyonaka, Osaka 560-8531, Japan}
\affiliation{Center for Quantum Information and Quantum Biology,
The University of Osaka, 1-2 Machikaneyama, Osaka 560-0043, Japan}
\affiliation{RIKEN Center for Quantum Computing (RQC), Hirosawa 2-1, Wako, Saitama 351-0198, Japan}

\date{\today}

\begin{abstract}

We propose a measurement-based fault-tolerant quantum computing (MB-FTQC) architecture for high-connectivity platforms, targeting early-FTQC devices with thousands to tens of thousands of physical qubits. Unlike earlier teleportation-based architectures that rely on code concatenation for scalability, MB-FTQC operates in a zero-concatenation regime with compact fixed-distance codes, keeping the qubit overhead within reach of such devices. MB-FTQC combines offline-verified logical resource states with Knill's error-correcting teleportation, allowing each online step to use single-round syndrome extraction and independent lookup-table decoding. Under circuit-level depolarizing noise, we study a Steane-code implementation with partially fault-tolerant analog \(R_Z(\theta)\) rotations and a Golay-code implementation with higher-order zero-level distillation for fault-tolerant \(T\)-state preparation. At a physical error rate of \(10^{-4}\), our estimates give a megaquop-scale equivalent \(T\)-count of \(1.89\times10^{6}\) for the Steane implementation and a gigaquop-scale characteristic \(T\)-count of \(1.87\times10^{9}\) for the Golay implementation. With sufficient parallel resource supply, MB-FTQC achieves a short online critical path, while resource estimates including verification failures and retries show competitive spacetime costs. These results show that compact, fixed-distance codes can support fault-tolerant computation at substantial scales in the early-FTQC regime, without code concatenation.

\end{abstract}

\maketitle

\section{Introduction}
\label{sec:introduction}
Reliable execution of large-scale quantum computations requires fault-tolerant quantum computing (FTQC) with quantum error correction (QEC).
Substantial experimental progress toward QEC has been made, including logical quantum memories~\cite{Ofek2016,Acharya2025} and fault-tolerant logical gates on multiple logical qubits~\cite{Bluvstein2024,paetznick2024,ryananderson2024}.
A central challenge is now to integrate these capabilities into practical FTQC architectures capable of sustaining long sequences of reliable logical operations.

Knill's teleportation-based QEC scheme~\cite{Knill2005,Knill2005Nature} has long provided an attractive building block for FTQC architectures. Fault-tolerantly prepared logical ancillas enable single-round syndrome extraction for online error correction, while offline resource-state preparation and verification can shorten the online critical path~\cite{Steane1997,Knill2005,Knill2005Nature,FY2010}. Architectures based on this approach have achieved high fault-tolerance thresholds under their respective noise models~\cite{Knill2005Nature,FY2010}.

Despite these advantages, their implementation has historically been challenging because they require nonlocal operations between encoded blocks, which are costly on hardware with local connectivity. This constraint is becoming less severe with the development of platforms offering high or reconfigurable connectivity, such as trapped ions and neutral-atom arrays. Recent experiments have demonstrated transversal logical gates and logical teleportation on such platforms~\cite{Bluvstein2024,ryananderson2024,bluvstein2025architecturalmechanismsuniversalfaulttolerant}. These advances therefore provide renewed motivation for revisiting teleportation-based FTQC architectures on emerging hardware.

The ability to reduce syndrome extraction to a single round is not unique to teleportation-based QEC schemes. Recent approaches combining transversal logical operations with correlated decoding provide an alternative route to single noisy syndrome-extraction rounds per logical operation, independent of code distance~\cite{cain2025fastcorrelateddecodingtransversal}. In such approaches, the number of syndrome-extraction rounds is reduced by exploiting error correlations across logical blocks and operations, which can make the classical decoding problem more global and circuit dependent. This motivates a complementary architectural question: can single-round error correction be retained while keeping the online decoding modular and localized to individual logical operations?

Several prominent teleportation-based FTQC architectures have achieved high fault-tolerance thresholds while relying on code concatenation for scalability. 
Examples include Knill's \(C_4/C_6\) architecture~\cite{Knill2005Nature} and the verified-cluster architecture of Fujii and Yamamoto~\cite{FY2010}.
Concatenation provides a systematic route to further suppress logical errors, but additional levels require more encoded blocks and increasingly hierarchical resource preparation. This overhead can be substantial for the early fault-tolerant devices with thousands to tens of thousands of physical qubits considered here. This raises a natural question: How far can the advantages of teleportation-based FTQC be retained without code concatenation?

To address these questions, we construct and evaluate measurement-based FTQC (MB-FTQC) in the \textit{zero-concatenation regime}, using compact fixed-distance codes without additional levels of concatenation. We encode logical operations into verified resource states and consume them through logical one-bit teleportation (LOBT)~\cite{ZLC2000,FY2010}, combining logical-gate execution and error correction in a common online primitive. In this sense, MB-FTQC realizes measurement-based computation~\cite{RB2001, RBB2003} at the logical-qubit level. Two successive LOBT steps are equivalent to Knill's error-correcting teleportation (ECT) gadget up to Pauli-frame updates~\cite{Knill2005,Knill2005Nature}. Resource-state preparation and verification are performed offline, keeping them off the online critical path. Each LOBT step is decoded independently using a small precomputed lookup table, keeping the online decoding modular and lightweight.

A closely related FTQC architecture is that of Bluvstein et al.~\cite{bluvstein2025architecturalmechanismsuniversalfaulttolerant}. The two works address complementary aspects of the same teleportation-centered paradigm: Bluvstein et al. experimentally demonstrate its key logical mechanisms and a universal fault-tolerant gate set, whereas MB-FTQC focuses on realizing and benchmarking this paradigm in a finite-code, zero-concatenation regime with explicit resource preparation and modular online decoding. Both architectures exploit high or reconfigurable connectivity, transversal logical operations, logical teleportation, and Pauli-frame tracking. The construction of Bluvstein et al. assumes high-quality logical \(\lvert T\rangle_L\) inputs and employs code- and circuit-tailored decoding, including lookup tables and correlated decoding. MB-FTQC instead explicitly constructs and benchmarks finite-code factories for non-Clifford resources and confines online decoding to independent, small lookup tables at each teleportation step.

We consider two implementations of MB-FTQC targeting different device scales.
For devices with thousands of physical qubits, we adopt the \([[7,1,3]]\) Steane code~\cite{Steane1996} with partially fault-tolerant analog \(R_Z(\theta)\) rotations~\cite{Akahoshi2024,ismail2025transversalstararchitecturemegaquopscale}.
For devices with tens of thousands of physical qubits, we use the \([[23,1,7]]\) Golay code~\cite{Steane1996PRA,Steane2003} and introduce \emph{higher-order zero-level distillation} (HO-ZLD), extending zero-level distillation~\cite{goto2016minimizing,ITHF2025} to provide \(T\)-state resources for a fully fault-tolerant Clifford+\(T\) implementation at the device scales considered here.

Our circuit-level simulations and resource estimates show that zero-concatenation MB-FTQC can reach substantial computational scales of practical interest in the early-FTQC regime. Under our circuit-level depolarizing noise model at \(p=10^{-4}\), the Steane implementation reaches the megaquop regime when the accuracy of its analog rotations is expressed as an equivalent \(T\)-count, while the Golay implementation reaches a characteristic \(T\)-count scale in the gigaquop regime based on the inverse extrapolated logical \(T\)-gate error rate.

We further find that these computational scales can be achieved with competitive resource costs. For Clifford operations, MB-FTQC has lower expected spacetime cost than a same-distance surface-code baseline with repeated syndrome extraction, while achieving an active volume comparable to that of transversal STAR with correlated decoding~\cite{ismail2025transversalstararchitecturemegaquopscale}. With sufficient parallel resource supply, it also provides the shortest online Clifford critical path among the schemes considered. For non-Clifford resource generation, HO-ZLD achieves lower estimated spacetime cost than fold-transversal cultivation~\cite{sahay2026foldtransversalcultivation} at comparable logical accuracy.
Overall, our MB-FTQC architecture combines teleportation-based error correction with verified resource-state preparation to provide a resource-efficient approach to early FTQC on high-connectivity platforms.

This paper is organized as follows.
Section~\ref{sec:preliminaries} reviews the background, and Section~\ref{sec:architecture} presents the architecture.
Sections~\ref{sec:simulation} and~\ref{sec:estimation} report simulations and resource estimates, respectively.
Section~\ref{sec:conclusion} concludes.

\section{Preliminaries}
\label{sec:preliminaries}
This section provides the minimal background for our architecture.
We first review logical one-bit teleportation (LOBT) and its connection to Knill's ECT gadget, followed by the verification of logical ancillas through entanglement purification.
We then summarize the STAR approach to analog rotations used in our Steane-code implementation.
Readers already familiar with these ingredients may proceed directly to Sec.~\ref{sec:architecture}.

\subsection{Logical one-bit teleportation and Knill's ECT}

LOBT serves as the fundamental building block of our fault-tolerant architecture. It generalizes the standard one-bit teleportation circuit~\cite{ZLC2000} by replacing each physical qubit with a \emph{code block} and performing all entangling and measurement operations transversally within a self-dual CSS code, such as the Steane or Golay code. By tracking Pauli byproducts in the Pauli frame, LOBT realizes logical gates by teleporting data through prepared logical ancillas while simultaneously extracting measurement outcomes as error-syndrome information~\cite{ZLC2000, FY2010}.

In the underlying physical one-bit teleportation, an input state $\ket{\psi}$ is entangled with $\ket{+}$ via a CZ gate and subsequently measured in the $X$ basis. The measurement outcome $m \in \{0,1\}$ determines the Pauli byproduct, yielding the output state $X^{m} H \ket{\psi}$~\cite{GC99}. 

$$
\includegraphics[width=0.3\textwidth]{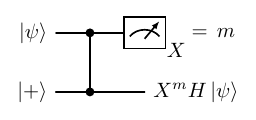}
$$

LOBT mirrors this circuit at the logical level: a logical $\ket{+}$ is prepared, a transversal CZ is applied between the data and ancilla blocks, and transversal $X$-basis measurements are performed on the data block. 
$$
\includegraphics[width=7cm]{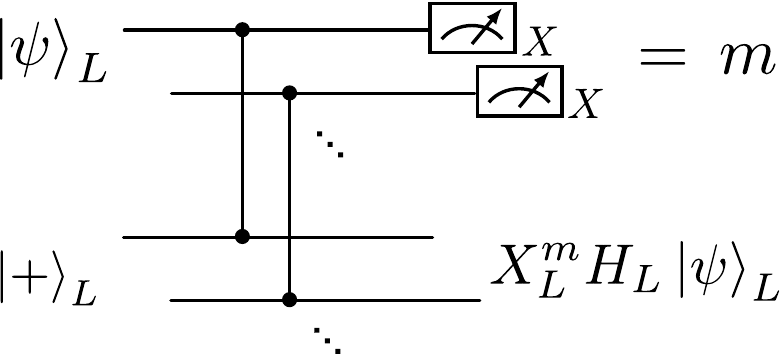}
$$
$Z$-type errors on the input block $\ket{\psi}_L$ can flip the parity of the raw logical $X$, obtained from the transversal $X$-basis measurement. The $X$-type syndromes obtained simultaneously from the same measurement detect these $Z$ errors. When the error weight is at most $\lfloor (d-1)/2 \rfloor$, a \emph{corrected} logical $X$-measurement bit $m$ can be computed from the syndrome information, allowing the corresponding logical byproduct $X_L^m$ to be determined. The resulting output state is $X_L^{m} H_L \ket{\psi}_L$.

Because $Z$ errors commute with CZ, they are confined to the input block and merely flip the measured syndrome (effectively \emph{absorbed} into the syndrome), preventing any propagation to the ancilla. This absorption provides built-in error correction.
$$
\includegraphics[width=5cm]{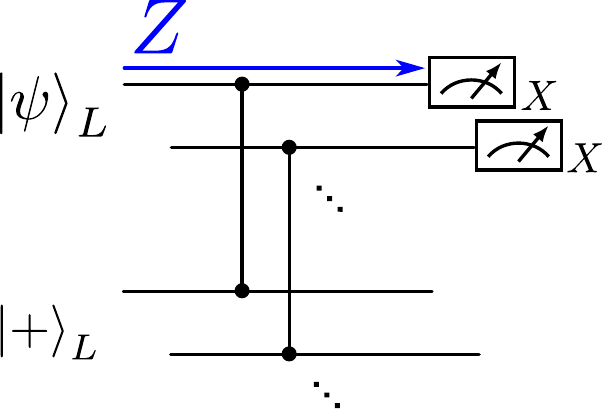}
$$

On the other hand, an $X$ error on the input block propagates through CZ into a $Z$ error on the output block ($\mathrm{CZ}\,(X\otimes I)\,\mathrm{CZ}=X\otimes Z$). By performing successive LOBT steps, these residual $Z$ errors are corrected in the next teleportation stage. Errors arising not only on the input state $|\psi\rangle_L$ but also within the LOBT subroutines (e.g., transversal CZ, ancilla block preparation) are handled in a similar way.
$$
\includegraphics[width=5cm]{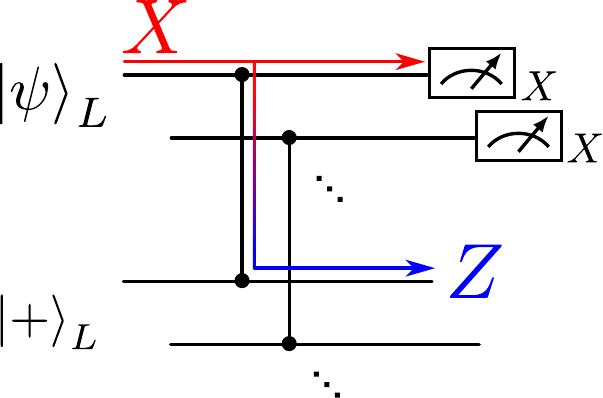}
$$
This limited propagation of physical errors across successive teleportation layers has also been observed experimentally in a neutral-atom processor (see Fig.~6 of Ref.~\cite{bluvstein2025architecturalmechanismsuniversalfaulttolerant}).

Since $H^2=I$, performing two LOBT steps in succession corrects both $X$ and $Z$ errors, effectively realizing the same transformation as Knill's ECT gadget, up to Pauli-frame updates~\cite{Knill2005, Knill2005Nature}.
Knill's gadget prepares a logical Bell pair and performs a joint Bell-basis measurement on the input logical qubit and one half of the Bell pair. A sequence of two LOBT circuits is equivalent to a single Knill's ECT gadget, with each LOBT step realizing half of the gadget. 
$$
\includegraphics[width=0.4\textwidth]{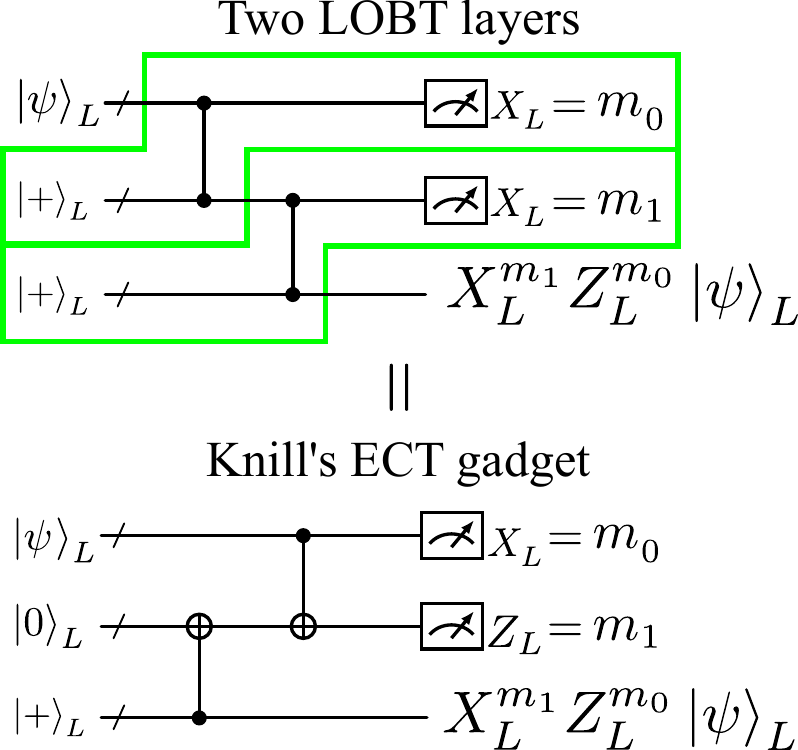}
$$

Many syndrome-extraction protocols, including Shor-style extraction~\cite{shor1997faulttolerantquantumcomputation} and flag-based schemes such as those in Ref.~\cite{Chao2018}, use repeated syndrome measurements to obtain reliable syndrome information.
In contrast, Knill's ECT gadget~\cite{Knill2005, Knill2005Nature}, and thus LOBT, permits \emph{single-shot} syndrome extraction, provided that fault-tolerantly prepared logical ancillas are supplied.

\subsection{Verification of logical ancillas by entanglement purification}
\label{subsec:method_zero_prep}

Single-shot ECT assumes verified logical ancillas. Because a single fault in a non-fault-tolerant encoder can produce a correlated intra-block error, we verify the encoded candidates using a two-round entanglement-purification protocol~\cite{Bennett1996, PR2012}:
\[
\includegraphics[width=0.4\textwidth]{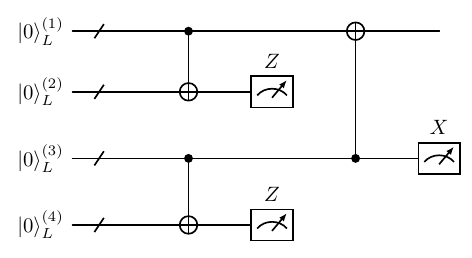}
\]
Four logical $\ket{0}_L$ candidates are produced using independent, non-fault-tolerant encoding circuits. The first round compares pairs of blocks using transversal CNOTs and qubit-wise $Z$-basis measurements. The measured $Z$-type stabilizers and logical-$Z$ eigenvalues detect $X$-type errors, including $Y$ errors. The two surviving blocks are then compared in the opposite CNOT direction, and one is measured qubit-wise in the $X$ basis. The resulting $X$-type stabilizers detect $Z$-type errors, including $Y$ errors; the logical $X$ eigenvalue is not checked because $\ket{0}_L$ is not its eigenstate. A nontrivial check causes the candidate to be discarded and the preparation to restart.

Although $Z$-type errors do not affect an immediate logical-$Z$ measurement, our prepared ancillas are used in subsequent gadgets, where residual $Z$-type errors can be transformed or propagated by Clifford operations and contribute to logical errors.
The second verification filters these residual $Z$-type errors before ancilla use.

The encoding circuits must be logically equivalent but physically distinct: using identical circuits for all four blocks can allow correlated errors to cancel during verification and escape detection~\cite{PR2012}. We therefore use code symmetries to permute physical-qubit indices between the encoding circuits. The code-specific choices used in MB-FTQC are given in Sec.~\ref{subsec:zero-implementation}.

\subsection{STAR-style analog rotations}

We briefly review the STAR architecture~\cite{Akahoshi2024}, which motivates the analog-rotation approach used in our Steane-code implementation (Sec.~\ref{subsec:method-analog-rot}). The STAR architecture combines error-corrected Clifford operations with non-fault-tolerant analog $R_Z(\theta)$ rotations, avoiding the cost of synthesizing arbitrary-angle rotations over Clifford+$T$ and distilling the required magic states.

The logical rotation is teleported using the ancilla $\ket{+_\theta}_L=R_Z(\theta)_L\ket{+}_L$. The measurement outcome determines whether $R_Z(\theta)_L$ or $R_Z(-\theta)_L$ is applied; in the latter case, a repeat-until-success (RUS) sequence uses ancillas $\ket{+_{2\theta}}_L,\ket{+_{4\theta}}_L,\ldots$ until the correction succeeds. This sequence requires two teleportation attempts on average. The residual analog-rotation error can be mitigated using probabilistic error cancellation~\cite{Akahoshi2024}.

\section{MB-FTQC Architecture}
\label{sec:architecture}

MB-FTQC is an architecture that realizes fault-tolerant quantum computation within the measurement-based paradigm~\cite{FY2010, FY2010verified}. The central idea is to pre-prepare logical ancilla states that correspond to the desired logical gates, and then apply those gates by executing LOBT between data blocks and the chosen ancilla blocks.

In our implementation, the syndrome and measurement outcomes from each LOBT step are processed independently using a precomputed lookup table for the chosen code. With fault-tolerantly prepared ancillas, the syndrome and raw logical measurement outcome from that step determine the corrected logical measurement bit and the corresponding Pauli-frame update. Correlated decoding that jointly processes syndromes across blocks or time steps is therefore not required. The lookup-table construction is detailed in Appendix~\ref{app:simulation-details-noise-model-and-decoding}.

To accommodate different device scales (i.e., the available number of physical qubits), we consider two implementation variants that differ in both the choice of error-correcting code and the realization of non-Clifford gates:

\begin{enumerate}
\item A 7-qubit Steane code implementing partially fault-tolerant analog logical $R_Z(\theta)$ rotations, targeting megaquop-scale (million-operation) quantum computation~\cite{Preskill_2025} with thousands of physical qubits.
\item A 23-qubit Golay code implementing fully fault-tolerant logical $T$ gates via magic-state distillation, targeting gigaquop-scale (billion-operation) quantum computation with tens of thousands of physical qubits.
\end{enumerate}

The remainder of this section describes logical ancilla preparation, logical gates through LOBT, and a functional three-zone organization for ancilla supply and physical-qubit allocation and reuse, which also serves as the basis for the subsequent resource estimates.

\begin{figure*}
    \centering
    \includegraphics[width=.7\linewidth]{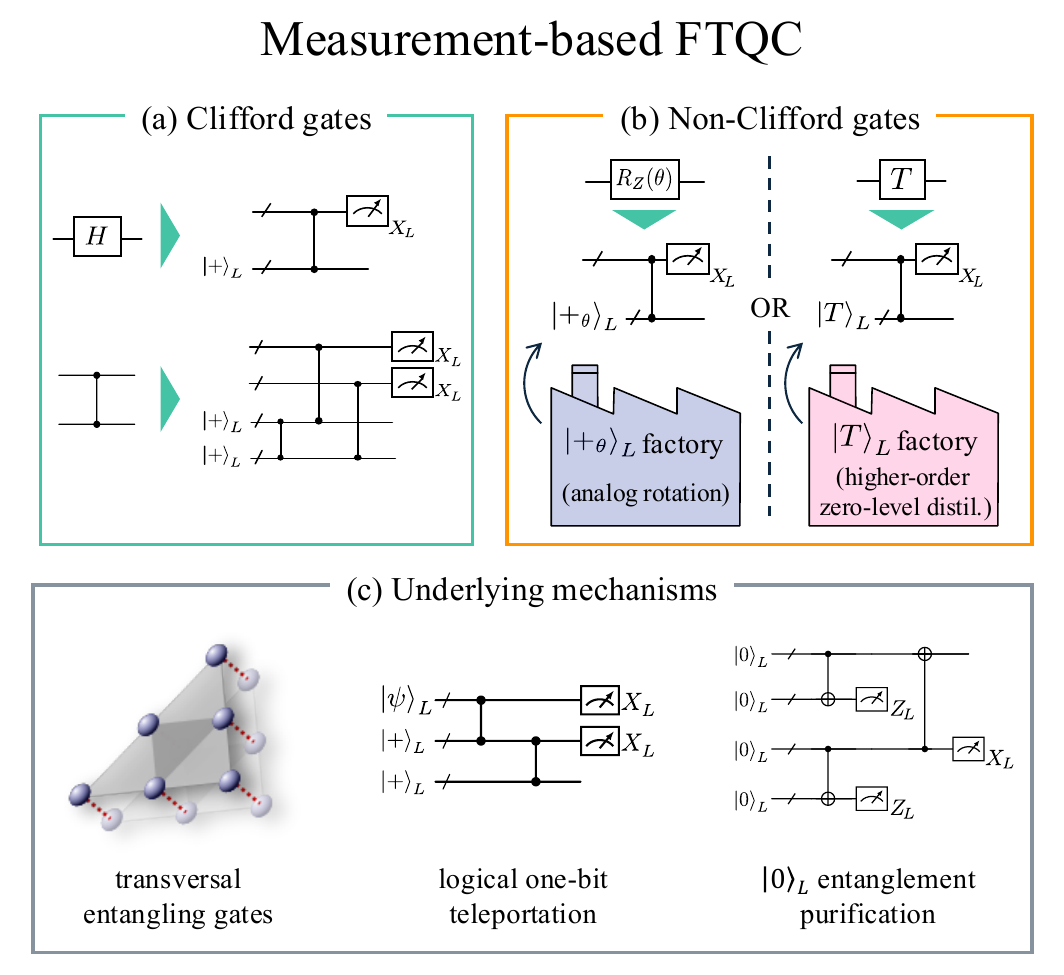}
    \caption{\textbf{Overview of MB-FTQC architecture.}
    (a) Clifford gates are executed by logical one-bit teleportation using pre-generated and verified logical ancillas; the gadget permits single-shot syndrome extraction, and compact lookup-table decoding determines the corrected logical measurement outcomes (Sec.~\ref{subsec:logical-h-cz}). (b) Two options are provided for non-Clifford gates: an analog $R_Z(\theta)$ using a $\lvert{+_\theta}\rangle_L$ factory (Sec.~\ref{subsec:method-analog-rot}), or a $T$ gate enabled by a $\lvert T\rangle_L$ factory based on higher-order zero-level distillation (Sec.~\ref{subsec:MagicT}). Prepared ancillas are applied with gate teleportation.
    (c) The architecture utilizes transversal entangling gates, logical one-bit teleportation (Sec.~\ref{sec:preliminaries}), and entanglement purification for $\lvert0\rangle_L$ state preparation (Sec.~\ref{subsec:method_zero_prep}).}
    \label{fig:overview}
\end{figure*}

\subsection{Code-specific logical $\ket{0}$ state preparation}
\label{subsec:zero-implementation}
For both implementation tracks, we instantiate the verification protocol of Sec.~\ref{subsec:method_zero_prep} with code-specific encoding circuits. For the Steane code, blocks 1 and 3 use the standard 8-CNOT encoder of Ref.~\cite{PR2012}, while blocks 2 and 4 use a code-symmetry permutation of that encoder.
Reusing the encoder within each pair is sufficient because, due to error degeneracy, any pure $Z$-type error on the Steane $\ket{0}_L$ state is equivalent to no error or a single-qubit $Z$ error~\cite{goto2016minimizing}.
The second verification is not required for fault tolerance, but we retain it to further suppress residual $Z$-type errors and reduce errors in subsequent gadgets using these ancillas.
For the Golay code, we use the four distinct encoders given in Table~4 of Ref.~\cite{PR2012}.
In this case, suppressing low-order correlated $Z$-type errors whose effective weight exceeds the number of faults is necessary for fault tolerance, and this is achieved by the second verification.
These choices preserve the target logical state while avoiding identical correlated-error paths across the compared blocks. The concrete Steane circuits and simulation procedures are detailed in Appendix~\ref{subsec:modeling}.

\subsection{Logical $H$ and CZ gates}
\label{subsec:logical-h-cz}

As outlined in Sec.~\ref{sec:preliminaries}, we implement logical gates via LOBT. The LOBT circuit for the logical Hadamard $H_L$ is given by the following circuit, which is equivalent to the physical-level circuit in Sec.~\ref{sec:preliminaries}:
\[
\includegraphics[width=0.40\textwidth]{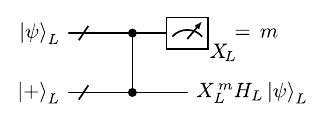}
\]
In this diagram, CZ and all measurements are executed transversally, i.e., qubit-wise on the underlying physical qubits of each code block. The measurement bit $m \in \{0, 1\}$ corresponds to the logical $X$ outcome which is corrected by $X$-type syndrome, obtained by the transversal $X$ measurements.

The LOBT circuit for the logical $\mathrm{CZ}$ is given by:
\[
\includegraphics[width=0.45\textwidth]{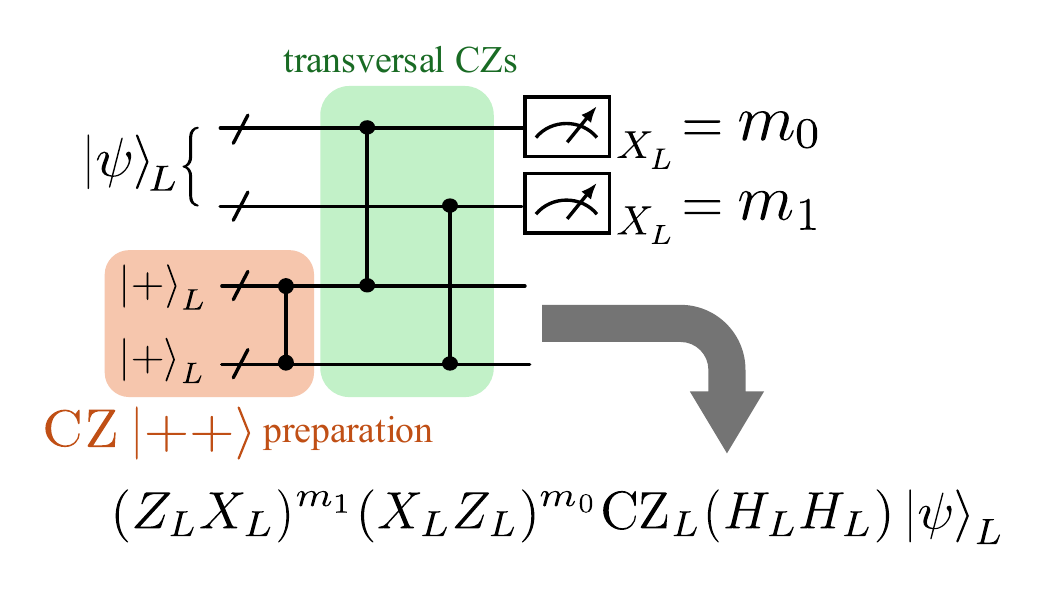}
\]
To apply a logical CZ between two logical qubits of the input state, we first prepare a two-qubit logical ancilla in the graph state
\[
|G_{ab}\rangle_L \;=\; \mathrm{CZ}_{L,ab}\,|+\rangle_{L,a}\,|+\rangle_{L,b},
\]
where $a$ and $b$ label the two ancilla blocks. Then teleport each data block individually, via separate LOBTs, into $a$ and $b$. Because the ancilla already carries the entangling CZ correlation, the net effect is a logical CZ between the original inputs, while the two LOBTs provide the same built-in error-correction behavior as in the single-qubit case. Note that this circuit also applies logical Hadamards to both input blocks in addition, right before the intended $\mathrm{CZ}_L$.

The following sections present implementations of logical non-Clifford gates. As outlined in Sec.~\ref{sec:preliminaries}, we offer two approaches whose choice depends on available qubit resources.

\subsection{Analog $R_Z$ rotation on the Steane code}
\label{subsec:method-analog-rot}
Following the STAR-style strategy reviewed in Sec.~\ref{sec:preliminaries}, we implement non-Clifford gates on the Steane code using a non-fault-tolerant analog $R_Z(\theta)$ rotation.
We pre-prepare the ancilla  $|+_\theta\rangle_L := R_Z(\theta)_L|+\rangle_L$ and then consume it on the data block using LOBT.

For the Steane code, we propose a preparation protocol for $|+_\theta\rangle_L$ that employs a newly designed shallow encoding circuit to minimize errors, combined with post-selection using Steane's syndrome extraction gadget~\cite{Steane1997}. Under a standard circuit-level noise model, the logical $Z$ error rate of the prepared ancilla is $p/15$ at leading order. Considering the repeat-until-success protocol for gate implementation, the logical $Z$ error rate of the $R_Z$ operation scales at leading order as $2p/15$.

\subsubsection{Preparation of the $\ket{+_\theta}_L$ ancilla}
\label{sec:ancilla_prep}
\begin{figure}[t!]
    \centering
    \includegraphics[width=1.1\linewidth]{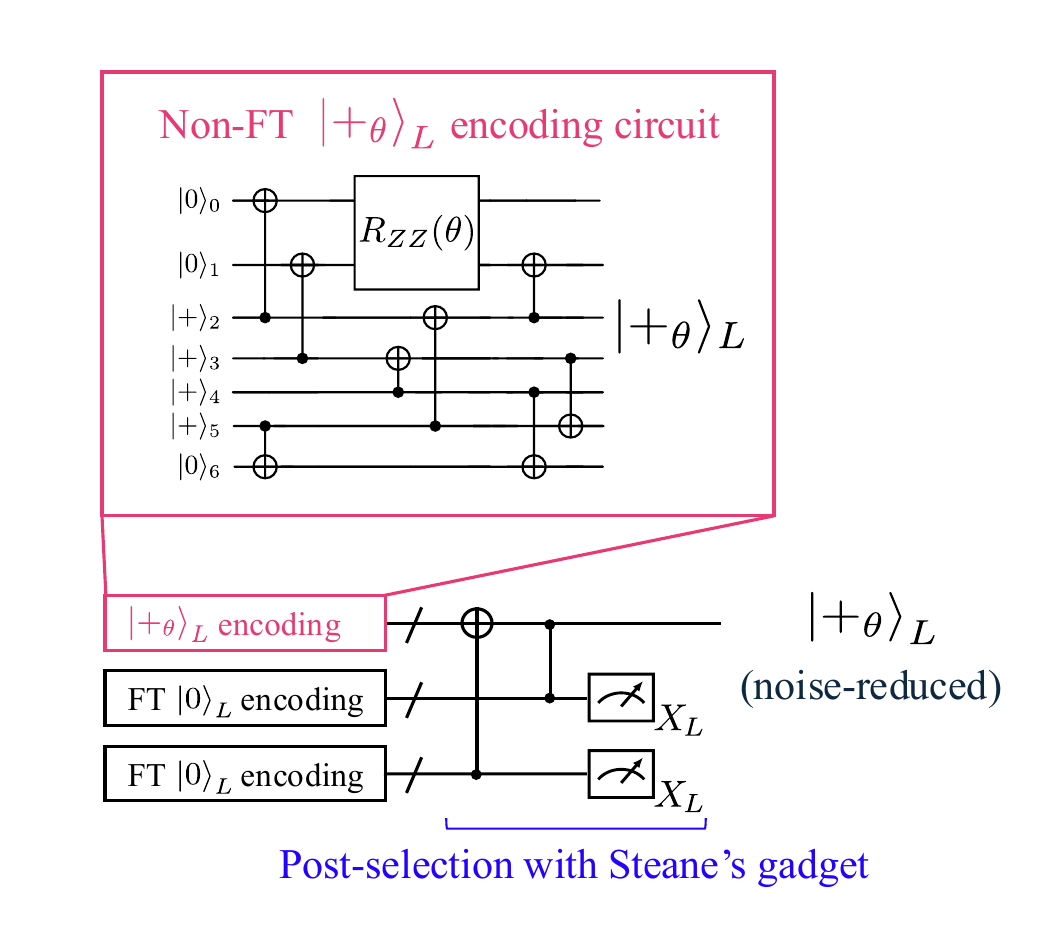}
    \caption{Full circuit for preparing the rotated Steane ancilla $\ket{+_\theta}_L$. (Top) A non-fault-tolerant encoding circuit using a single analog two-qubit rotation $R_{ZZ}(\theta)$. (Bottom) Post-selection with Steane's gadget that couples the prepared block to two fault-tolerant $\ket{0}_L$ ancillas and extracts both $Z$- and $X$-type syndromes; only runs with trivial syndromes are accepted.}
    \label{fig:analog_rotation_circuit}
\end{figure}

We employ a two-step protocol that yields a leading-order logical $Z$-error rate of approximately $p/15$ under a standard circuit-level noise model.
The preparation procedure can be divided into the following two parts:

\paragraph*{{\rm (a)} Encoding of non-FT $\ket{+_\theta}_L$ state.} 
The tailored encoding circuit shown in Fig.~\ref{fig:analog_rotation_circuit} (top) employs only nine two-qubit gates, including a single analog $ZZ$ rotation, $R_{ZZ}(\theta)=e^{-i\theta Z\otimes Z/2}$. 
Compared with a conventional Steane encoding circuit (e.g., Fig.~2 of Ref.~\cite{ITHF2025}), the reduced two-qubit count lowers both temporal cost and physical error exposure.

\paragraph*{{\rm (b)} Post-selection using Steane's gadget.} 
Because the prepared $\ket{+_\theta}_L$ ancilla is noisy, we apply Steane's gadget by coupling it to two fault-tolerantly prepared $\ket{0}_L$ blocks to measure all $X$- and $Z$-type syndromes (see Fig.~\ref{fig:analog_rotation_circuit}, bottom).
Trials yielding any non-trivial syndrome are discarded, thereby removing all first- and second-order faults that anticommute with some stabilizers of the code.

All correlated errors from two-qubit gates in the preparation circuit are detected, except for the Pauli $ZZ$ error in the two-qubit depolarizing channel at the $R_{ZZ}$ rotation.
Since the Pauli $ZZ$ at that stage is equivalent to a logical $Z$ and thus indistinguishable from the intended logical $Z$ rotation, the method effectively reaches the fundamental lower bound of the logical error rate under our circuit‑level depolarizing noise model.
Compared with conventional Steane encoding circuits that allow arbitrary single-qubit state injection, this approach provides a substantial reduction in the logical error rate. 
The detailed justification for the logical error rate of $p/15$ achieved by this procedure is provided in Appendix~\ref{appsec:analog-rot}.

\subsubsection{Applying $R_Z(\theta)_L$ via LOBT and a repeat-until-success protocol}

We implement $R_Z(\theta)_L$ using LOBT with the ancilla $\ket{+_\theta}_L$ with the following LOBT circuit:
\[
\includegraphics[width=0.40\textwidth]{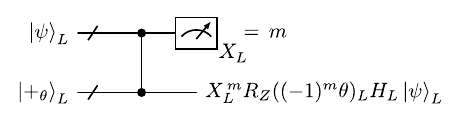}
\]

A single LOBT attempt yields either $R_Z(\theta)_L$ or $R_Z(-\theta)_L$, as indicated by the measurement outcome. We use the RUS sequence reviewed in Sec.~\ref{sec:preliminaries}, whose expected number of attempts is two. Appendix~\ref{app:rus-latency} analyzes the resulting latency distribution in a multi-qubit circuit. Under a conservative synchronized-barrier model, the expected latency of a layer containing 100 simultaneous RUS rotations is approximately $7.98$ RUS rounds, about four times the mean latency of an isolated rotation.

Given the ancilla's leading-order logical $Z$ error $p/15$ and an average of two attempts, the error contribution per logical $R_Z(\theta)$ rotation is approximately $2p/15$ (to first order in $p$).
It should also be noted that in a single LOBT attempt for $R_Z$, both $R_Z$ and $H_L$ are applied. Consequently, during the RUS procedure, an $H_L$ gate must be inserted between successive LOBT attempts to cancel this effect.

These error rate scalings are based on the uniform depolarizing noise model considered here. Under more general stochastic Pauli noise, the leading residual logical \(Z\) error is set by the \(ZZ\) component of the \(R_{ZZ}\) gate that passes post-selection, and can therefore be larger for \(ZZ\)-biased noise. Since this residual error forms a stochastic logical \(Z\) channel, it can be mitigated by probabilistic error cancellation as in the STAR architecture~\cite{Akahoshi2024}, at the cost of increased sampling overhead. In contrast, coherent over-rotation in \(R_{ZZ}\) is generally not removed by post-selection. If such errors dominate, \(R_{ZZ}\) can instead be decomposed into two CNOTs and a virtual-\(Z\) rotation, avoiding coherent over-rotation from the native analog \(R_{ZZ}\) gate when the CNOTs are sufficiently well calibrated.

\subsection{$T$ gate via higher-order zero-level distillation on the Golay code}
\label{subsec:MagicT}
Since the Golay code leads to \(O(p^{4})\) logical error scaling, it is natural to implement the entire Clifford+$T$ gate set in a fully fault-tolerant manner. We therefore use magic-state distillation to fault-tolerantly prepare high-fidelity \(\ket{T}_L= R_Z(\pi/4)_L \ket{+}_L\) states, and apply the logical \(T\) gate via gate teleportation with transversal operations, as in the previous section.
The dominant failure mode applies \(T_L^{\dagger}\) instead; this is corrected deterministically by a logical $S$ gate (itself transversal for self-dual CSS codes such as the Steane and Golay codes), since
\(
S_L T_L^{\dagger} = e^{-i\pi/4} T_L.
\)

To reduce the cost of $\ket{T}_L$ preparation, we propose \emph{higher-order zero-level distillation}, a low physical-qubit overhead distillation protocol. 
This scheme extends zero-level distillation~\cite{goto2016minimizing, ITHF2025}, which operates directly on noisy physical gates without concatenation, by repeating a verification block consisting of a Hadamard test followed by stabilizer post-selection using Steane's gadget.
Here, ``higher-order'' refers to the higher-order suppression of the output logical error obtained through repeated verification.
While conceptually related to a protocol developed for a two-dimensional color code~\cite{chamberland2020very}, our approach is specifically tailored for high-connectivity hardware, employing the Golay code and using one physical control ancilla and one flag qubit in the Hadamard test.

The procedure of higher-order zero-level distillation for preparing a fault-tolerant $\ket{T}_L$ state proceeds as follows:
\begin{enumerate}
  \item Initialize the data block in $\ket{T}_L$ non-fault-tolerantly.
  \item Perform a Hadamard test of $H_{XY}\coloneqq(X+Y)/\sqrt{2}$ using transversal Clifford gates and the identity $T X T^{\dagger} = H_{XY}$.
  \item Apply Steane's gadget for post-selection and discard on detection.
  \item Repeat steps (ii)-(iii) for \(r\) rounds to raise the detected error order.
  \item Output the accepted state as \(\ket{T}_L\), which is to be used by gate teleportation.
\end{enumerate}

Since \(T\ket{+}\) is the \(+1\) eigenstate of \(H_{XY}\), we prepare \(T\ket{+}\) by repeatedly performing a Hadamard test of \(H_{XY}\), followed by post-selection using Steane's gadget (see Fig.~\ref{fig:golay-distillation}). Using the identity \(T X T^{\dagger}=H_{XY}\), together with the transversal Clifford gates of the Golay code, the Hadamard test can be implemented as shown in Fig.~\ref{fig:golay-distillation-hadamard-test-implementation}. The additional ancilla in the Bell pair is a flag qubit that detects single \(X\) or \(Y\) errors on the control ancilla during the multi-target CNOT sequence, which could otherwise produce correlated errors on multiple data qubits. Runs with a nontrivial flag outcome are rejected. By preventing these dangerous correlated errors arising from a single fault from being accepted undetected, the flag qubit contributes to the fault tolerance of the protocol. The ancillas used in Steane's gadget are logical \(\ket{0}\) states prepared via entanglement purification in the same manner as in the analog-rotation case.

We repeat this verification block \(r\) times, with \(r\le 3=(d-1)/2\) for the distance-\(7\) Golay code. For \(r=3\), state initialization and the three verification blocks form four stages. Thus, if at most three faults occur, the pigeonhole principle guarantees that at least one stage is fault free. Once a fault-free stage restores the correct logical magic state, the subsequent flag and error-detection steps suppress errors arising from the remaining faults. Our circuit-level noise simulations show the expected $O(p^{r+1})$ scaling, including $O(p^4)$ for $r=3$, providing numerical confirmation of the expected fault-tolerant behavior. The practical choice of \(r\) is discussed later in Sec.~\ref{sec:simulation}.

\begin{figure}[t!]
\centering
\[
\Qcircuit @C=.5em @R=.7em {
\lstick{}           &      &     &     & \push{\ket{+}} & \ctrl{1}                & \gate{H}  & \meter \\
\lstick{T\ket{+}_L} & /\qw & \qw & \qw & \qw            & \gate{H_{XY}}           & \gate{PS} & \qw    & \qw & \qw \gategroup{1}{4}{3}{8}{.6em}{-} \\
\lstick{}           &      &     &     &                &                         &           & \\
\lstick{}           &      &     &     &                &                         &           & \\
\lstick{}           &      &     &     &                & \mbox{Repeat $r$ times} & 
}
\]
\caption{The higher-order zero-level distillation protocol. $PS$ stands for post-selection using Steane's gadget.}
\label{fig:golay-distillation}
\end{figure}

\begin{figure}[t!]
\centering
\[
\Qcircuit @C=.5em @R=.6em {
\lstick{\ket{0}} & \targ            & \qw      & \qw      & \qw      & \targ     & \qw                      & \meter \\
\lstick{\ket{+}} & \ctrl{-1}        & \ctrl{1} & \qw      & \ctrl{3} & \ctrl{-1} & \gate{H}                 & \meter \\
\lstick{}        & \gate{T^\dagger} & \targ    & \qw      & \qw      & \gate{T}  & \multigate{2}{PS} & \qw \\
\lstick{}        & \cdots           &          & \cdots   &          & \cdots    &                   &     \\
\lstick{}        & \gate{T^\dagger} & \qw      & \qw      & \targ    & \gate{T}  & \ghost{PS}        & \qw \\
}
\]

\caption{The implementation of the Hadamard test used in the distillation protocol (Fig.~\ref{fig:golay-distillation}).
         The top two qubits are the flag qubit (initialized in \(\ket{0}\)) and the control ancilla (initialized in \(\ket{+}\)), respectively. The remaining qubits form a logical qubit encoded in the Golay code.}
\label{fig:golay-distillation-hadamard-test-implementation}
\end{figure}

\subsection{Functional three-zone organization}
\label{subsec:functional-organization}

\begin{figure}[t!]
    \centering
    \includegraphics[width=1.\linewidth]{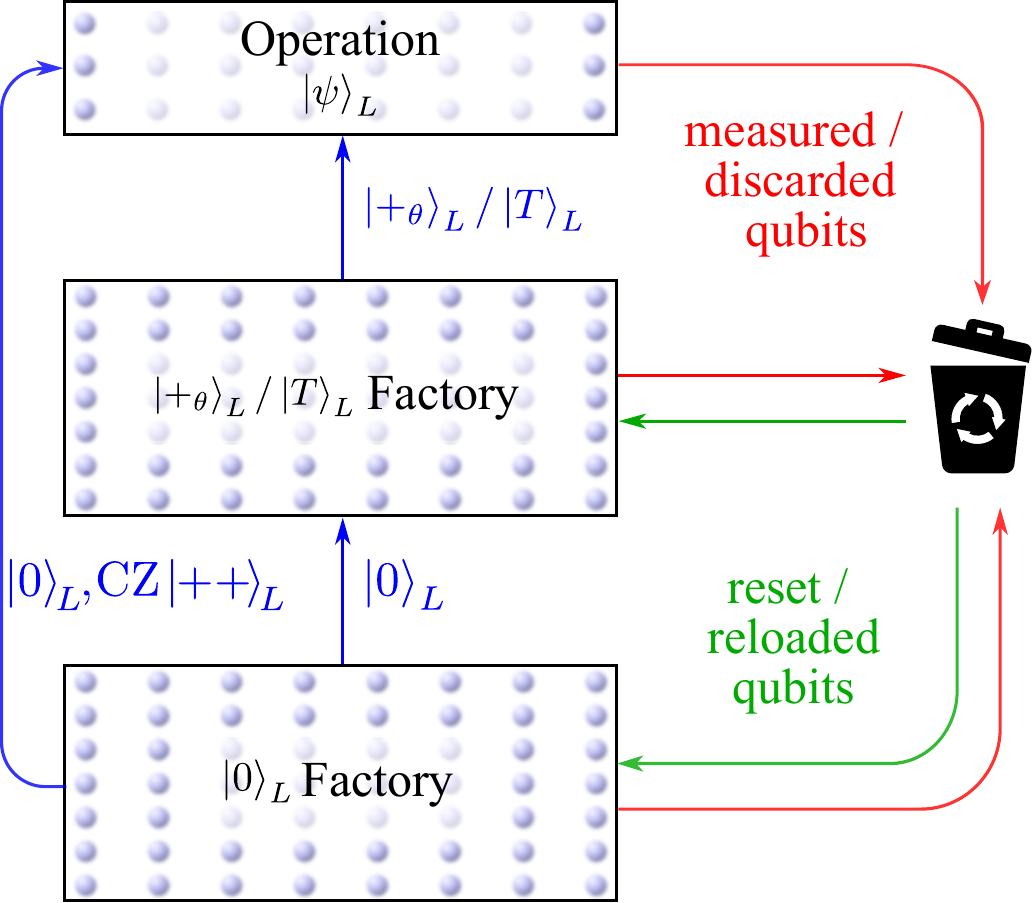}
    \caption{Functional three-zone organization of MB-FTQC. Blue arrows indicate the supply of logical ancillas. Red arrows indicate measured or discarded qubits, and green arrows indicate reset or reloaded qubits supplied to the factories. The vertical arrangement is schematic and does not prescribe a device geometry.}
    \label{fig:zones}
\end{figure}

To organize the execution of logical circuits in MB-FTQC, we introduce a high-level three-zone model (Fig.~\ref{fig:zones}). This model specifies architectural choices that are not determined by individual logical-gate circuits alone: where ancillas are prepared and supplied, how measured or rejected physical qubits are reused, and how physical qubits are allocated to these stages. It also provides a reference for the resource estimates in Sec.~\ref{sec:estimation}. The zones are distinguished by their functional roles rather than by requiring distinct hardware capabilities:

\begin{enumerate}[label=(\roman*),wide=0pt,itemsep=0pt,parsep=0pt]
\item $\ket{0}_L$ \textbf{Factory Zone}: logical zero ancillas are prepared by entanglement purification and supplied to the other factory or the Operation Zone. Ancillary $\mathrm{CZ}\ket{++}_L$ states are also prepared here for logical CZ gates.

\item $\ket{+_\theta}_L / \ket{T}_L$ \textbf{Factory Zone}: non-Clifford resource states are prepared using the protocols described above, consuming $\ket{0}_L$ ancillas supplied by the other factory.

\item \textbf{Operation Zone}: logical gates on the data qubits are implemented through LOBT using ancillas supplied by the factories.
\end{enumerate}

We assume that all three zones support physical single- and two-qubit Clifford operations and single-qubit measurements in the computational basis. Measurements in the required bases can be implemented using Clifford basis changes. The hardware provides the connectivity required by the encoding, verification, and transversal inter-block circuits. The corresponding non-Clifford resource-state factory additionally supports the physical operations required by its preparation protocol: $R_{ZZ}(\theta)$ for the Steane implementation and $T$ and $T^\dagger$ for the Golay implementation.

Qubits can be coherently transported between zones to supply the prepared ancillas. The factories are supplied with physical qubits initialized to $\ket{0}$ through reset or reloading. Ancilla preparation can proceed in parallel with operations on the data qubits. We also assume classical processing of measurement outcomes for post-selection, decoding, Pauli-frame updates, and the required feedforward.

\section{Numerical simulation}
\label{sec:simulation}

This section benchmarks logical state preparation and a universal gate set for the Steane- and Golay-code implementations. The resulting logical error rates are used in the performance and resource estimates of Sec.~\ref{sec:estimation}.

\subsection{Simulation strategy and scope}

We use the Clifford simulator Stim~\cite{gidney2021stim}. Logical $\ket{0}$ preparation and the Clifford $H$ and CZ gates are simulated directly. The non-Clifford contributions are instead estimated using separate Clifford proxy circuits for the analog $R_Z(\theta)$ rotation and the logical $T$-state preparation, respectively, allowing both to be simulated with Stim.

For the Steane analog rotation, we use a Clifford proxy circuit to isolate the leading \(O(p)\) logical-\(Z\) contribution after post-selection. The only non-Clifford operation in the preparation circuit is \(R_{ZZ}(\theta)\), which is followed in our noise model by an angle-independent two-qubit Pauli channel. Because the subsequent post-selection tests code stabilizers, whether a given Pauli fault is detected is also independent of \(\theta\).
Error-propagation analysis shows that all first-order faults are detected except the \(ZZ\) fault at the \(R_{ZZ}\) location, which leaves the measured syndromes unchanged and acts as a logical \(Z\). Logical \(X\) errors, by contrast, appear only at \(O(p^2)\). We can therefore set \(\theta=0\) while retaining the same two-qubit noise channel. Since \(R_{ZZ}(0)=I\), the ideal output becomes \(\ket{+}_L\) and the entire circuit is Clifford. An accepted logical \(Z\) error maps \(\ket{+}_L\) to \(\ket{-}_L\), so its probability can be extracted directly from the logical \(X\) measurement outcome in Stim.
The teleportation contribution is simulated separately, and the mean RUS repetition factor is included when reporting the logical \(R_Z(\theta)_L\) error.

For Golay higher-order zero-level distillation, we replace $T$ by $S$ and simulate preparation of the proxy state $\ket{S}_L\coloneqq S_L\ket{+}_L$. Following Ref.~\cite{gidney2024}, we estimate the target-state preparation error as $p_{\ket{T}}^{\rm est}\approx 2p_{\ket{S}}$. This factor of two applies only to the ancilla-preparation contribution; the independently simulated teleportation error is added separately.

State-preparation logical error rates quantify deviations from the target logical state after ideal decoding, conditioned on passing post-selection.
For logical Clifford gates, we estimate the total probability $p_{\mathrm{tot}}=\sum_{P\ne I}p_P$ of nonidentity logical Pauli errors in the effective channel following the ideal gate.
In the gate simulations, residual physical errors on the prepared ancillas are included using the effective error model described in Appendix~\ref{subsec:modeling}.
We simulate repeated gate applications to account for the accumulation of physical errors on the data.

The resulting logical error rates are obtained directly from Monte Carlo simulations unless otherwise stated. Proxy-based estimates and extrapolated values are explicitly identified where used. In particular, the Golay results at $p=10^{-4}$ are extrapolated where direct sampling is infeasible, as noted with the corresponding results below.

We use a uniform circuit-level depolarizing model in which state preparation,
one- and two-qubit gates, and measurement have the common error scale $p$.
Although SPAM errors can exceed gate errors on some platforms, including
superconducting devices, representative trapped-ion and neutral-atom
experiments report SPAM or readout errors comparable to or smaller than
two-qubit-gate errors~\cite{Ransford2026,Evered2023}. The complete channel
definitions, fault locations, and decoder construction are given in
Appendix~\ref{app:simulation-details}.

We omit idling (memory) errors in our main simulations. For
representative trapped-ion and neutral-atom devices, the memory error
accumulated over a two-qubit gate timescale is estimated to satisfy
$p_{\mathrm{idle}}/p_{2Q}<10^{-2}$, and is therefore much smaller than
the dominant two-qubit gate error
(Appendix~\ref{app:memory-errors}). We further verified this
approximation by explicitly including memory errors in the Steane-code
encoding simulations. Even for the deliberately pessimistic choice \(p_{\mathrm{idle}}=p\), where \(p\) denotes the common physical-error scale, the logical error rate remains essentially unchanged within statistical uncertainty. The main effect is instead a modest increase in the post-selection overhead (Appendix~\ref{app:memory-error-sensitivity}). We therefore neglect
memory errors in the main simulations, as their inclusion does not
affect our qualitative conclusions.
\subsection{Logical $\ket{0}$ State Preparation}
\label{subsec:zero-prep-results}

We benchmark the code-specific implementations defined in
Sec.~\ref{subsec:zero-implementation}. The encoding circuits, post-selection
criteria, and logical-error evaluation procedure are detailed in
Appendix~\ref{app:simulation-details}.

\begin{figure}[htbp]
  \centering
  \includegraphics[width=0.9\linewidth]{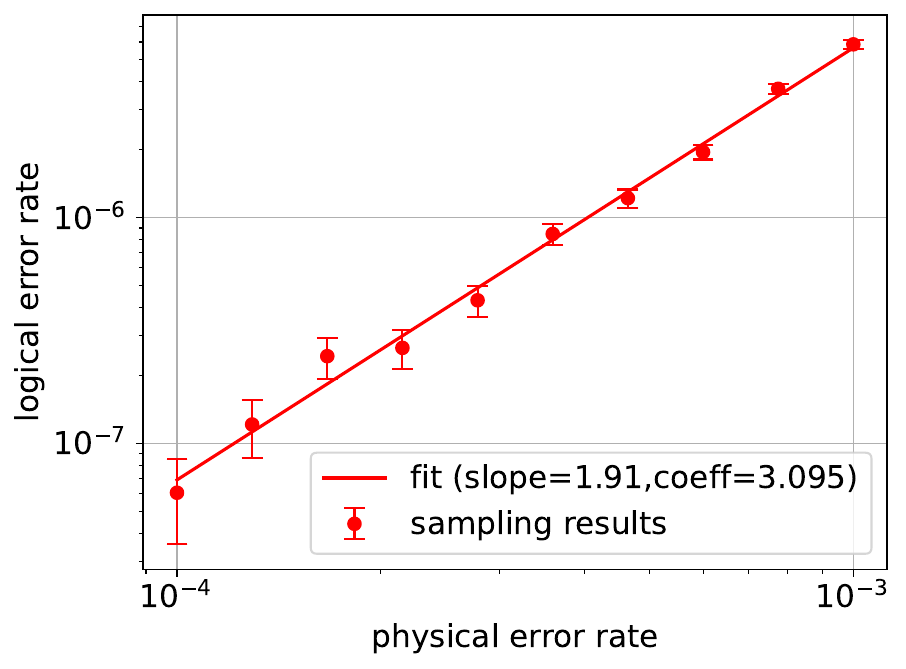}\\[0.5cm]
  \includegraphics[width=0.9\linewidth]{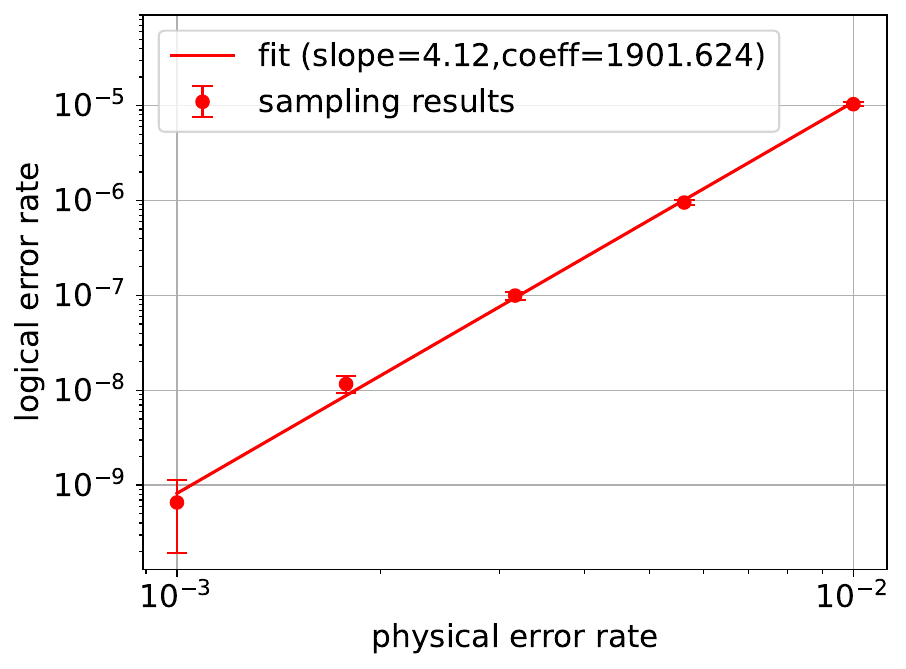}
  \caption{Logical error rates of prepared $\ket{0}_L$ state as a function of the physical error rate $p$. (top) Steane code. (bottom) Golay code. The numbers of samples are $10^8$ and $5\times 10^9$, respectively.}
  \label{fig:zero-prep-error}
\end{figure}

\begin{figure}[htbp]
  \centering
  \includegraphics[width=0.9\linewidth]{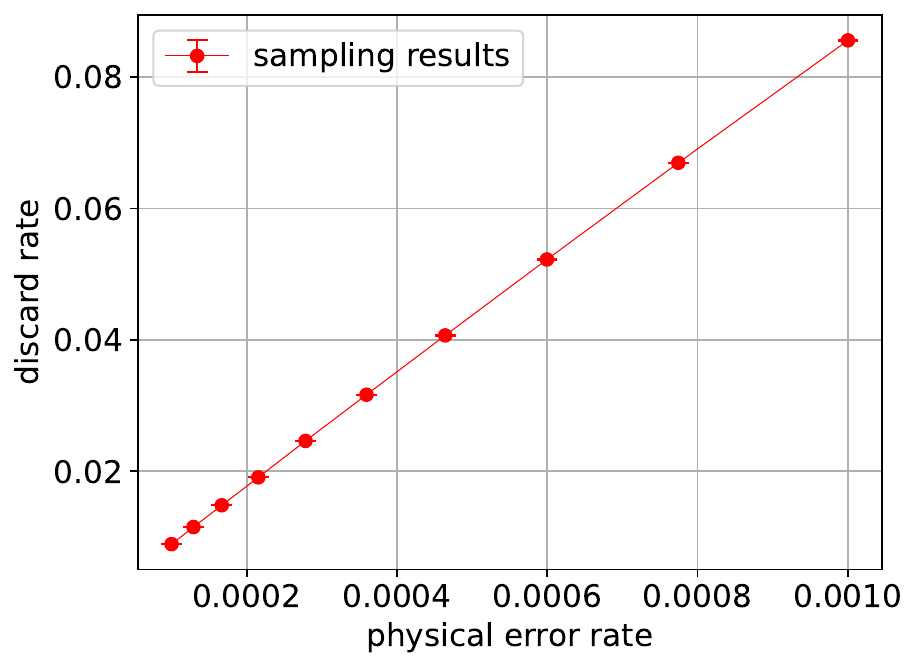}\\[0.5cm]
  \includegraphics[width=0.9\linewidth]{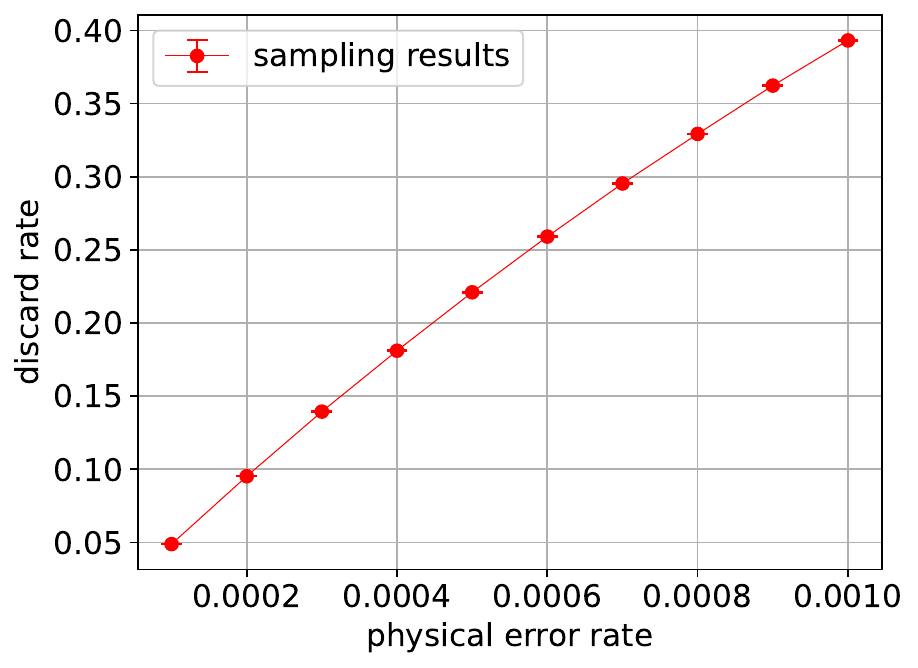}
  \caption{Discard rate of $\ket{0}_L$ state purification as a function of the physical error rate $p$. (top) Steane code. (bottom) Golay code. The numbers of samples are $10^8$, $10^7$, respectively.}
  \label{fig:zero-prep-discard}
\end{figure}

The resulting logical error rates are shown in Fig.~\ref{fig:zero-prep-error}, where the data are fitted to curves of the form $C p^{\alpha}$. As expected, the Steane and Golay codes exhibit $O(p^2)$ and $O(p^4)$ scaling, respectively. 
Figure~\ref{fig:zero-prep-discard} shows the post-selection discard rates. 
At a physical error rate of $p = 10^{-3}$, the discard rate is below 10\% for the Steane code and below 50\% for the Golay code, so that retrying until success is a realistic option.
At $p = 10^{-4}$, it further decreases to below 1\% (Steane) and 5\% (Golay), while the corresponding logical error rates become sufficiently low to enable computations of meaningful scale.
The complete retry and physical-operation costs of this purification, including its use as an input to higher-level resource-state preparations, are quantified in Table~\ref{tab:postselection-overhead}.

\subsection{Logical Clifford Gate Error Rates}

Following Ref.~\cite{Steane2003}, we evaluate the logical $H$ and CZ gates
using repeated-application benchmarks that retain physical errors accumulated
between successive operations. We report estimates of upper bounds on the
total nonidentity logical Pauli error probabilities. The effective ancilla
model, the error-rate definitions, and the observed transient and periodic behavior are
described in Appendix~\ref{appsubsec:repeated-clifford}.

\begin{figure}
    \centering
    \includegraphics[width=0.9\linewidth]{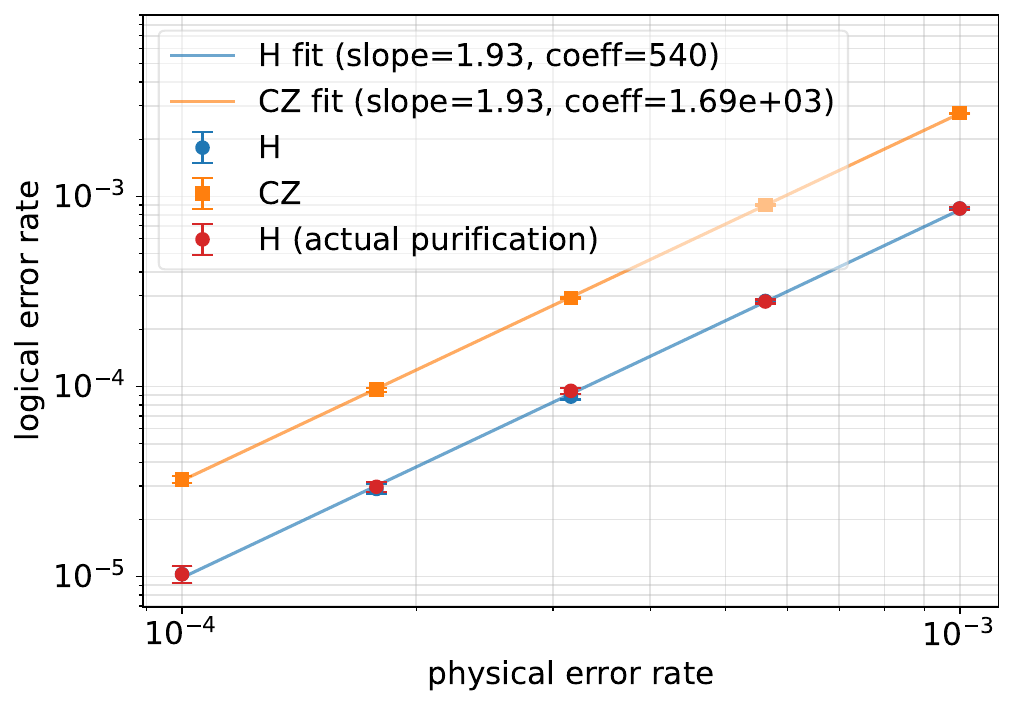}

    \includegraphics[width=0.9\linewidth]{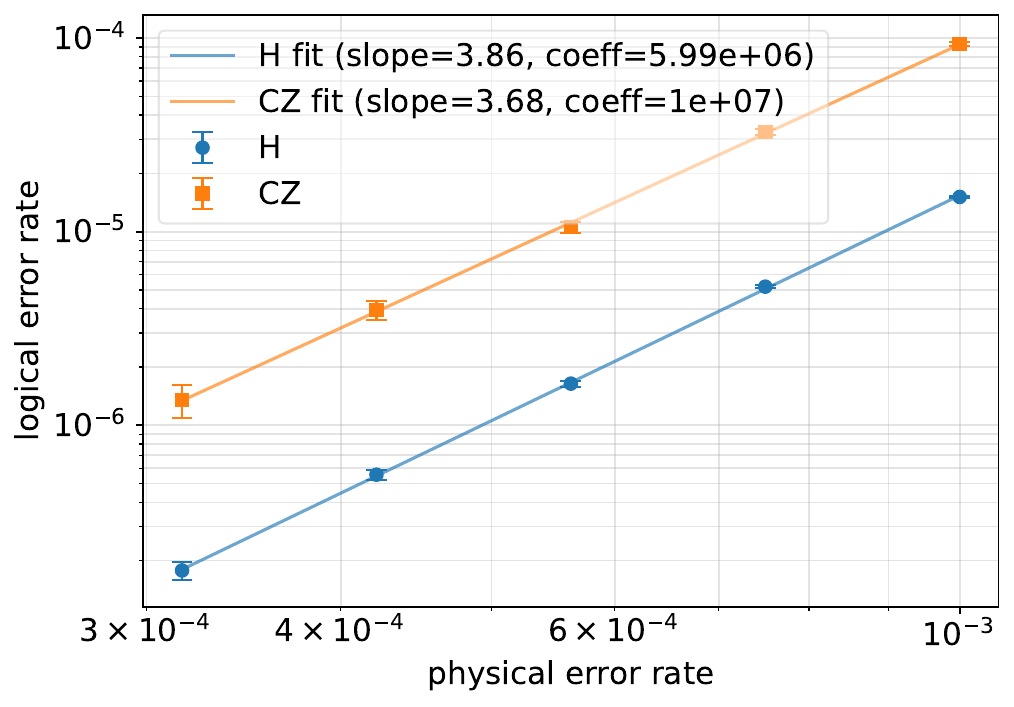}
    \caption{Estimated upper bounds on the total nonidentity logical Pauli error probabilities for $H$ and CZ gates in the Steane (top) and Golay (bottom) codes, using the error distribution model on purified $\ket{0}_L$ ancillas. The numbers of samples are $10^7$ (Steane, $H$ and CZ), $5\times 10^8$ (Golay, $H$), and $10^7$ (Golay, CZ), respectively.  For $H$ gates of the Steane code, we also plot the result using those ancillas prepared by explicit purification without the model (red circles).}
    \label{fig:Clifford}
\end{figure}

As shown in Fig.~\ref{fig:Clifford}, these estimates exhibit the expected
scaling: $O(p^2)$ for the Steane code and $O(p^4)$ for the Golay code. For the
Steane $H$ gate, simulations using explicitly purified ancillas agree with the
effective-model results, validating its use in the more computationally
demanding simulations.

\subsection{Analog $R_Z$ rotation on the Steane code}
\label{sec:simulation-rz}

\begin{figure}
    \centering
    \includegraphics[width=0.9\linewidth]{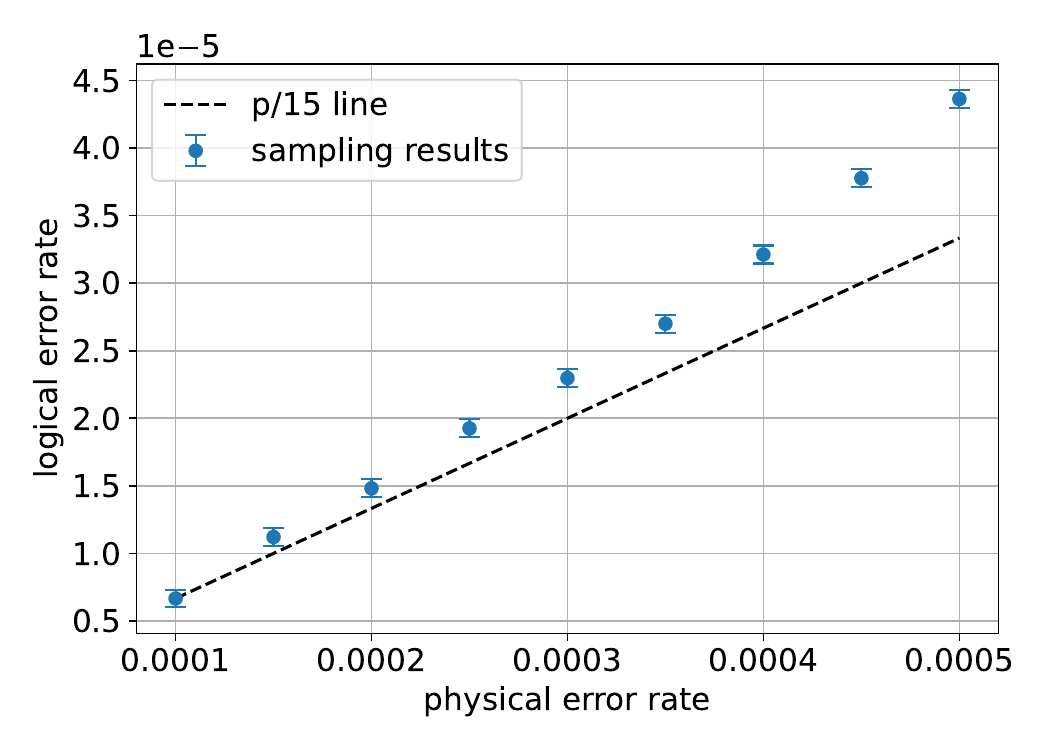}
    \includegraphics[width=0.9\linewidth]{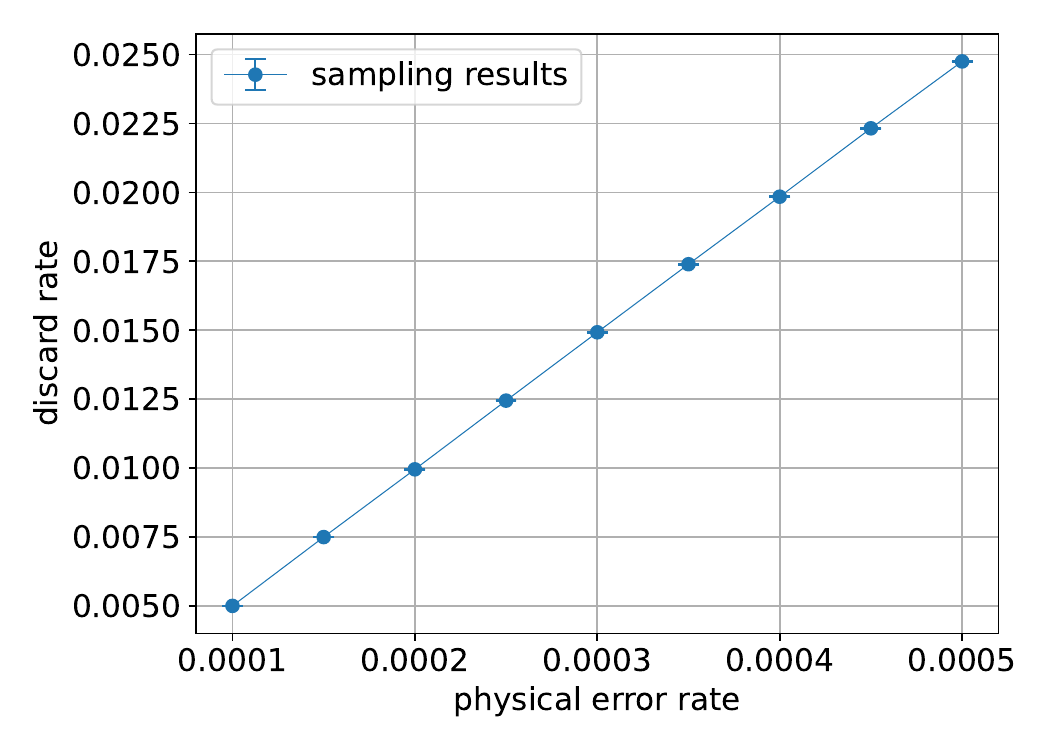}
    \caption{Stim simulation of the $\theta=0$ proxy circuit for Steane analog-rotation ancilla preparation. (top) Accepted logical-$Z$ error rate, whose leading-order behavior is $p/15$, used to estimate the $\ket{+_\theta}_L$ preparation error. (bottom) Post-selection discard rate. The number of samples is $10^8$.}
    \label{fig:rz-prep}
\end{figure}

We evaluate the ancilla-preparation and teleportation contributions separately.
Under our circuit-level noise model, the $\theta=0$ proxy simulation gives an
accepted logical-$Z$ error probability $p_{\mathrm{rot}}$ with leading-order
behavior $p_{\mathrm{rot}}\approx p/15$
(Fig.~\ref{fig:rz-prep}). Error-propagation analysis shows that the logical-$X$
contribution starts only at $O(p^2)$. For the independently simulated
teleportation contribution, we use an upper-bound estimate
$p_{\mathrm{tele}}$, which scales as $O(p^2)$.
The proxy-circuit construction and error-estimation details are given in Appendix~\ref{app:simulation-details}.

Including the mean of two RUS attempts, the error per logical $R_Z(\theta)_L$ rotation is
\[
p_{R_Z} \approx 2\bigl(p_{\mathrm{rot}} + p_{\mathrm{tele}}\bigr),
\]
where $p_{\mathrm{tele}}=O(p^2)$.

For $p=10^{-4}$, Table~\ref{tab:steane-rz-error-breakdown} gives
$p_{\mathrm{rot}}=6.670\times10^{-6}$ and
$p_{\mathrm{tele}}=5.520\times10^{-6}$, yielding an estimated total logical
rotation error of $p_{R_Z}=2.438\times10^{-5}$.
This gives a rotation-count estimate of
$1/p_{R_Z}\approx4.10\times10^4$.
Using Clifford+$T$ synthesis~\cite{Akahoshi2024} at the corresponding target
accuracy $\delta=p_{R_Z}$, each analog rotation is equivalent to
$3\log_2(1/\delta)\approx46$ $T$ gates. The total
equivalent $T$-count is therefore approximately $1.89\times10^6$.
Therefore, at \(p = 10^{-4}\), MB-FTQC with the Steane code and analog rotation enables megaquop-scale computation without the need for concatenation.

\subsection{$T$ gate via higher-order zero-level distillation on the Golay code}

We evaluate the protocol using the $\ket{S}_L$ proxy described above. Its
logical error rate is converted to the estimated $\ket{T}_L$ preparation error
using $p_{\ket{T}}^{\rm est}\approx2p_{\ket{S}}$, while the gate-teleportation
contribution is evaluated separately. The effective model for the verified
$\ket{0}_L$ inputs and the teleportation benchmark are detailed in
Appendix~\ref{app:simulation-details}.

\begin{figure}[t!]
\centering
\includegraphics[width=1.05\linewidth]{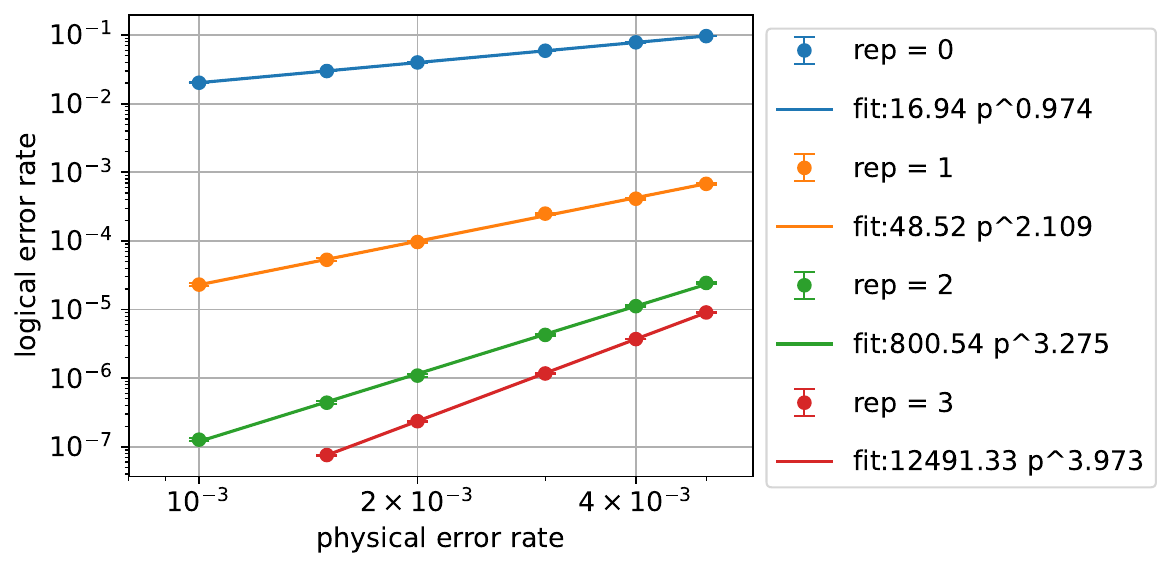}
\caption{Logical error probability of the accepted $\ket{S}_L$ proxy state. The target-state preparation error is estimated as $p_{\ket{T}}^{\rm est}\approx2p_{\ket{S}}$. Here $r$ is the number of verification repetitions. The number of samples varies with the physical error rate, up to $2\times10^{12}$ per data point.}
\label{fig:msd-error-rate}
\end{figure}

\begin{figure}[t!]
\centering
\includegraphics[width=0.9\linewidth]{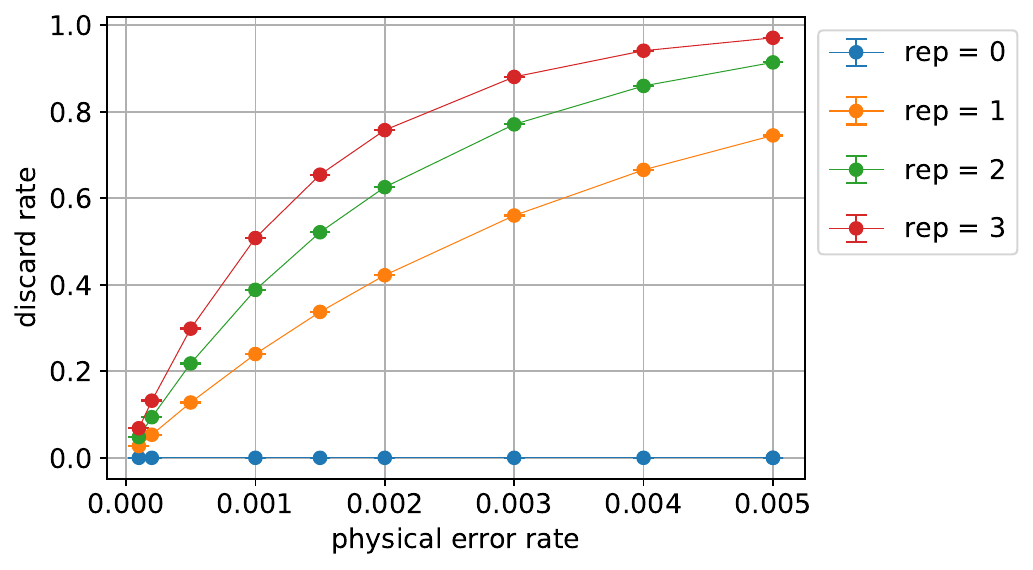}
\caption{Discard rate of the $\ket{S}_L$ proxy distillation circuit. Here $r$ is the number of verification repetitions. The number of samples varies with $r$ and the physical error rate, up to $2\times10^{12}$ per data point.}
\label{fig:msd-success-rate}
\end{figure}

Figures~\ref{fig:msd-error-rate} and~\ref{fig:msd-success-rate} show the logical error and discard rates of the $\ket{S}_L$ proxy circuit, respectively. Here $r$ denotes the repetition count for the Hadamard test and Steane's gadget.
We focus on the physical-error range \(10^{-4} \leq p \leq 10^{-3}\).  The corresponding acceptance probabilities and complete physical-operation costs, including the preparation of verified input ancillas, are summarized in Table~\ref{tab:postselection-overhead}.

The logical error probability scales as $O(p^{r+1})$.
To match the $O(p^4)$ logical error probability offered by the Golay code, one would set $r=3$.
However, since this distillation is an error-detection protocol, whose constant prefactor is generally smaller than that of error correction, $r<3$ can be sufficient in some practical situations.
Our data show that at \(p \approx 10^{-3}\), \(r=2\) already achieves a logical error rate lower than that of error-corrected Clifford operations, as shown in Table~\ref{tab:logical-error-rates-1e-3}, leaving little motivation to use \(r=3\); at smaller \(p\), \(r=3\) becomes beneficial.

The estimated error per logical $T$ gate is the sum of the proxy-based ancilla-preparation estimate and the independently simulated upper-bound estimate for the transversal gate-teleportation contribution,
\[
p_T = p_{\ket{T}}^{\rm est}+p_{\mathrm{tele}}
\approx 2 p_{\ket{S}} + p_{\mathrm{tele}},
\]
where $p_{\ket{S}}$ is the logical error rate of the accepted $\ket{S}_L$ proxy state. The numerical contributions are listed in Table~\ref{tab:golay-t-error-breakdown}.

For the Golay code, direct sampling at $p = 10^{-4}$ is infeasible due to the extremely low logical error rate. 
For these estimates, we extrapolate using fits with the exponent fixed to four, $p_L=Cp^4$.
The fitted coefficients give $p_{\ket{S}} = 1.450\times10^4p^4$ (for $r=3$)
and $p_{\mathrm{tele}}=5.331\times10^6p^4$, leading to
$p_T=5.360\times10^{-10}$ at $p=10^{-4}$.
The inverse $1/p_T\approx1.87\times10^9$ defines a characteristic $T$-count scale in the gigaquop regime, corresponding to an estimated mean of one logical $T$-gate error.

\vspace{5mm}

Tables~\ref{tab:logical-error-rates-1e-3} and \ref{tab:logical-error-rates} summarize the estimated logical gate error rates at physical error rates of $p = 10^{-3}$ and $p = 10^{-4}$, respectively, for both implementation tracks. The $H$ and CZ entries are estimates of upper bounds on the total nonidentity logical Pauli error probabilities, as defined in Appendix~\ref{appsubsec:repeated-clifford}.
For the Golay code, the results in the latter table are fitted and extrapolated to $p = 10^{-4}$ as described above.

\begin{table}[t]
    \centering
    \caption{Estimated logical gate error rates for $p=10^{-3}$. The $H$ and CZ entries are upper-bound estimates of the total nonidentity logical Pauli error probabilities.}
    \label{tab:logical-error-rates-1e-3}
    \setlength{\tabcolsep}{4pt}
    \renewcommand{\arraystretch}{1.2}
    \begin{tabular}{@{}lccc@{}}
        \toprule
        & $H$ & CZ & non-Clifford \\
        \midrule
        Steane & $8.641\times10^{-4}$ & $2.740\times10^{-3}$
        & \shortstack{$1.184\times10^{-3}$  $(R_Z)$} \\
        Golay & $1.512\times10^{-5}$ & $9.339\times10^{-5}$
        & \shortstack{$5.584\times10^{-6}$  $(T,\,r=2)$} \\
        \bottomrule
    \end{tabular}
\end{table}

\begin{table}[t]
    \centering
    \caption{Estimated logical gate error rates for $p=10^{-4}$. The $H$ and CZ entries are upper-bound estimates of the total nonidentity logical Pauli error probabilities.}
    \label{tab:logical-error-rates}
    \setlength{\tabcolsep}{4pt}
    \renewcommand{\arraystretch}{1.2}
    \begin{tabular}{@{}lccc@{}}
        \toprule
        & $H$ & CZ & non-Clifford \\
        \midrule
        Steane & $1.028\times10^{-5}$ & $3.241\times10^{-5}$
        & \shortstack{$2.438\times10^{-5}$ $(R_Z)$} \\
        Golay & $1.525\times10^{-9}$ & $9.443\times10^{-9}$
        & \shortstack{$5.360\times10^{-10}$ $(T,\,r=3)$} \\
        \bottomrule
    \end{tabular}
\end{table}

\section{Performance and Resource Estimation}
\label{sec:estimation}

In this section, we first quantify the post-selection overhead of the resource-state factories. We then estimate the spatial resources (number of physical qubits) required by the proposed architecture and assess its computational performance, using an error-budget-based quantum-volume (QV) proxy. Finally, under the same high-connectivity assumption, we compare its spacetime costs with surface-code-based baselines for both Clifford operations and non-Clifford resource-state generation.
Our resource estimates use a circuit-level implementation model under the hardware assumptions described in Sec.~\ref{subsec:functional-organization} and do not separately model device-specific overheads from transport, routing, or qubit reloading.

\subsection{Post-selection Overhead}
\label{subsec:postselection-overhead}

The resource states are prepared offline and each preparation circuit is repeated independently until it is accepted. Table~\ref{tab:postselection-overhead} reports both the resulting mean repetition counts and the average physical-operation cost per accepted output. We count each physical state preparation, one-qubit gate, two-qubit gate, and measurement as one physical operation. For simplicity and to obtain a conservative upper estimate of the operation cost, even when rejection can be determined partway through an attempt, we count the attempt as if all remaining operations in the preparation circuit were executed; no reduction from early termination is included.

For logical zero preparation, which does not consume verified logical input ancillas, the average cost is
\begin{align*}
C_{\rm avg}(\lvert 0\rangle_L)
&=
\frac{C_{\rm direct}(\lvert 0\rangle_L)}
{P_{\rm acc}(\lvert 0\rangle_L)},\\
C_{\rm avg}(\lvert\psi\rangle_L)
&=
\frac{C_{\rm direct}(\lvert\psi\rangle_L)
+n_0 C_{\rm avg}(\lvert0\rangle_L)}
{P_{\rm acc}(\lvert\psi\rangle_L)}.
\end{align*}
Here, \(C_{\rm direct}\) denotes the physical-operation count of a single execution of the corresponding preparation circuit, excluding the cost of any verified input ancillas, and \(n_0\) is the number of verified \(\lvert0\rangle_L\) ancillas consumed per attempt. 
The second expression applies to the composite resource states considered here, namely \(\lvert\psi\rangle_L=\lvert+_\theta\rangle_L\) for the Steane code and \(\lvert\psi\rangle_L=\lvert T\rangle_L\) for the Golay code. Their preparation circuits consume \(n_0=2\) and \(n_0=2r\) verified \(\lvert0\rangle_L\) states per attempt, respectively, where \(r=2\) or \(3\) for the Golay protocol. The direct cost \(C_{\rm direct}(\lvert\psi\rangle_L)\) excludes the cost of preparing these input ancillas.
Thus, \(C_{\rm avg}\) recursively includes retries within the \(\lvert0\rangle_L\) factories, rather than only repetitions of the listed composite-state preparation circuit.

\begin{table*}[t]
\centering
\caption{
Post-selection overheads for resource-state preparation.
Here \(P_{\rm acc}\) is the acceptance probability of the listed preparation circuit and \(\langle N_{\rm att}\rangle=1/P_{\rm acc}\) is its mean number of repetitions. For preparations consuming verified \(\lvert0\rangle_L\) ancillas, these two quantities are conditioned on the required ancillas being available, whereas \(C_{\rm avg}\) recursively includes all operations and retries required to prepare them.
For simplicity, a rejected attempt is charged the full circuit cost even if rejection could be determined before the circuit is completed. By not crediting any reduction from early termination, this convention gives a conservative upper estimate of the average operation cost.
}
\label{tab:postselection-overhead}
\setlength{\tabcolsep}{8pt}
\renewcommand{\arraystretch}{1.15}
\begin{tabular}{lccc ccc}
\toprule
& \multicolumn{3}{c}{\(p=10^{-3}\)} & \multicolumn{3}{c}{\(p=10^{-4}\)} \\
\cmidruleleft
\cmidruleright
Resource
& \(P_{\rm acc}\) & \(\langle N_{\rm att}\rangle\) & \(C_{\rm avg}\)
& \(P_{\rm acc}\) & \(\langle N_{\rm att}\rangle\) & \(C_{\rm avg}\) \\
\midrule
Steane \(\lvert0\rangle_L\)
& 0.9144 & 1.094 & 132
& 0.9911 & 1.009 & 122 \\
Steane \(\lvert+_\theta\rangle_L\)
& 0.9511 & 1.051 & 343
& 0.9950 & 1.005 & 308 \\
Golay \(\lvert0\rangle_L\)
& 0.6068 & 1.648 & 997
& 0.9512 & 1.051 & 636 \\
Golay \(\lvert T\rangle_L\), \(r=2\)
& 0.6124 & 1.633 & 7,377
& 0.9521 & 1.050 & 3,229 \\
Golay \(\lvert T\rangle_L\), \(r=3\)
& 0.4920 & 2.033 & 13,673
& 0.9316 & 1.073 & 4,896 \\
\bottomrule
\end{tabular}
\end{table*}

The acceptance probabilities remain substantial at both physical error rates. Even at \(p=10^{-3}\), the mean number of repetitions of a listed preparation circuit is at most \(2.033\); at \(p=10^{-4}\), it is at most \(1.073\). The complete nested cost is captured by \(C_{\rm avg}\). For example, at \(p=10^{-3}\), the \(r=3\) Golay protocol consumes on average \(12.20\) verified Golay \(\lvert0\rangle_L\) states per accepted \(\lvert T\rangle_L\) state. Preparing these inputs requires \(20.10\) executions of the underlying \(\lvert0\rangle_L\) preparation circuit, including retries, leading to a total of \(C_{\rm avg}=13{,}673\) physical operations per accepted output.

\subsection{Space Overhead}

We estimate the physical-qubit footprint of a reference allocation within the functional three-zone organization. Let $\eta = N_\mathrm{phys} / N_{\mathrm{data}}$ denote the number of physical qubits required per data logical qubit in our architecture.
As a baseline, we assume a simple \emph{per-qubit} layout in which each data logical qubit is associated with two ancilla factories, one for $\ket{0}_L$ and another for $\ket{+_\theta}_L$ or $\ket{T}_L$, so that non-Clifford operations (e.g., $R_Z(\theta)$ or $T$) can be performed by preparing the corresponding ancilla and consuming it in an operation zone via gate teleportation.
Each factory has a fixed physical-qubit allocation and retries failed preparations independently until success. Operations wait for the required ancillas when necessary. These retries increase preparation latency and the operation count quantified in Sec.~\ref{subsec:postselection-overhead}, but do not change the fixed physical-qubit footprint estimated below.
Spacetime-cost estimates including these retries are presented in
Sec.~\ref{subsec:surface-code-resource-comparison}.
Likewise, random RUS outcomes increase runtime and spacetime volume but do not change this fixed footprint; their circuit-level latency is analyzed in Appendix~\ref{app:rus-latency}.
Two-qubit Clifford gates such as CZ are implemented by preparing a purified pair of $\ket{0}_L$ states and entangling them into a two-qubit graph state using $H$ and CZ operations within the $\ket{0}_L$ factory. This process fits within the same per-qubit factory layout.

\subsubsection{Steane code with analog rotation}
\begin{figure}[t]
    \centering
    \includegraphics[width=1.\linewidth]{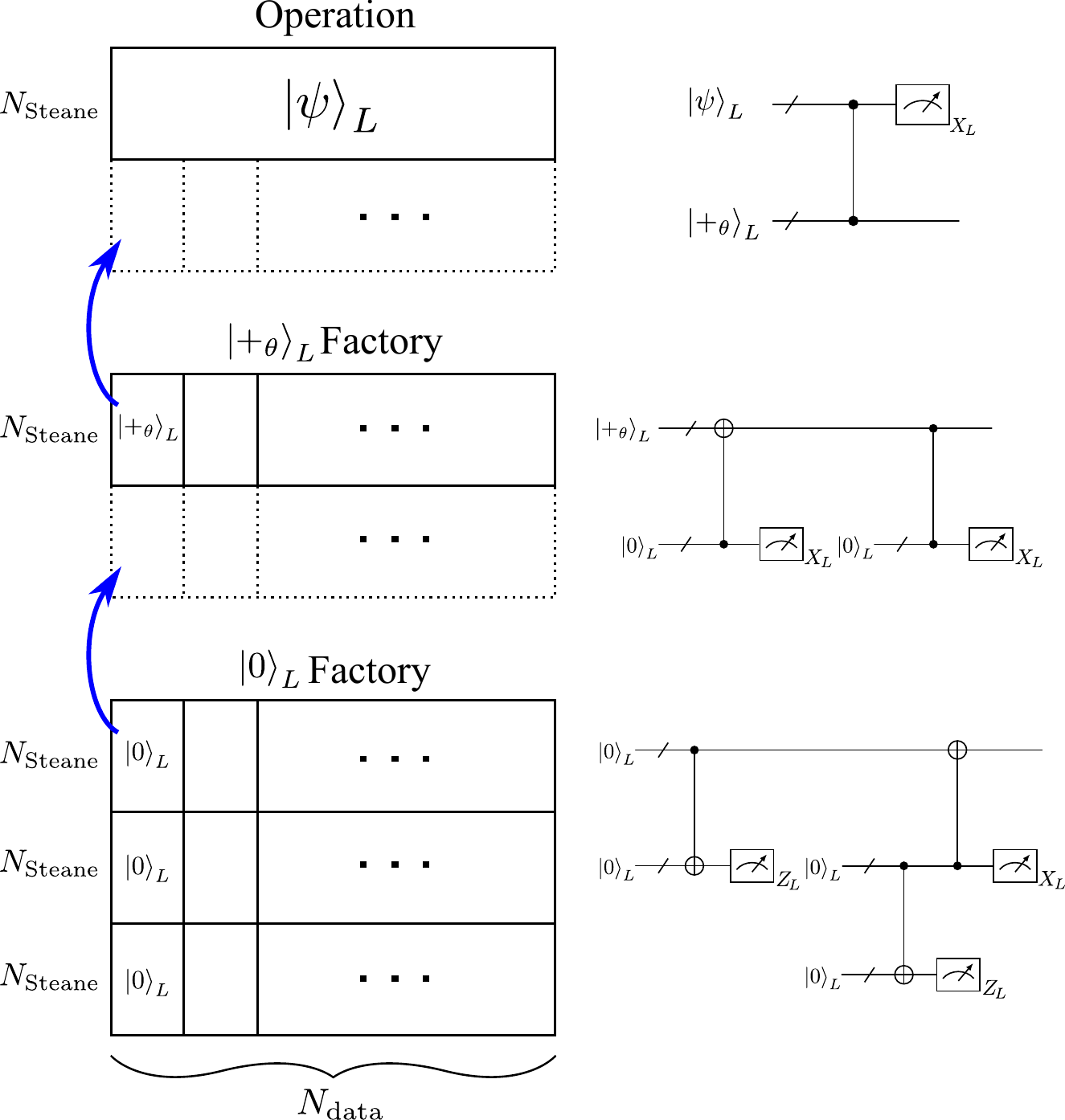}
    \caption{Schematic qubit allocation within the functional three-zone organization
    of MB-FTQC for executing a circuit on $N_{\mathrm{data}}$ logical qubits
    with analog rotations. Dashed lines indicate regions overlapping with
    preceding areas, which are excluded from the resource estimation.}
    \label{fig:analog-factory}
\end{figure}

Figure~\ref{fig:analog-factory} shows the qubit allocation based on the functional three-zone organization discussed in Sec.~\ref{sec:architecture}. Including state preparation, the layout for each data logical qubit consists of five code blocks in total: three allocated to the $\ket{0}_L$ factory, one to the $\ket{+_\theta}_L$ factory, and one serving as the operation zone. The width (number of columns) corresponds to the number of data logical qubits, $N_{\mathrm{data}}$.
As shown in the right side of the figure, the circuits for the $\ket{0}_L$ and $\ket{+_\theta}_L$ factories are deformed to fit within three and two code blocks, respectively. In the $\ket{+_\theta}_L$ factory and the operation zone, dashed lines indicate placeholder areas that overlap with the preceding factories and are therefore excluded from the qubit budget. 
Since the Steane code encodes one logical qubit into $N_{\mathrm{Steane}}=7$ physical qubits per block, the total physical-qubit budget per data logical qubit is
\[
\eta_{\mathrm{Steane}} = 7 \times 5 = 35 .
\]

\subsubsection{Golay code with magic-state distillation}
Proceeding analogously for the Golay code, we again require five code blocks per data logical qubit, and we add two extra physical qubits in the $T$-factory for the Hadamard test ancillas. As each Golay code block uses $N_{\mathrm{Golay}}=23$ physical qubits, the resulting estimate is
\[
\eta_{\mathrm{Golay}} = 23 \times 5 + 2 = 117 .
\]

\begin{figure}[t]
    \centering
    \includegraphics[width=1\linewidth]{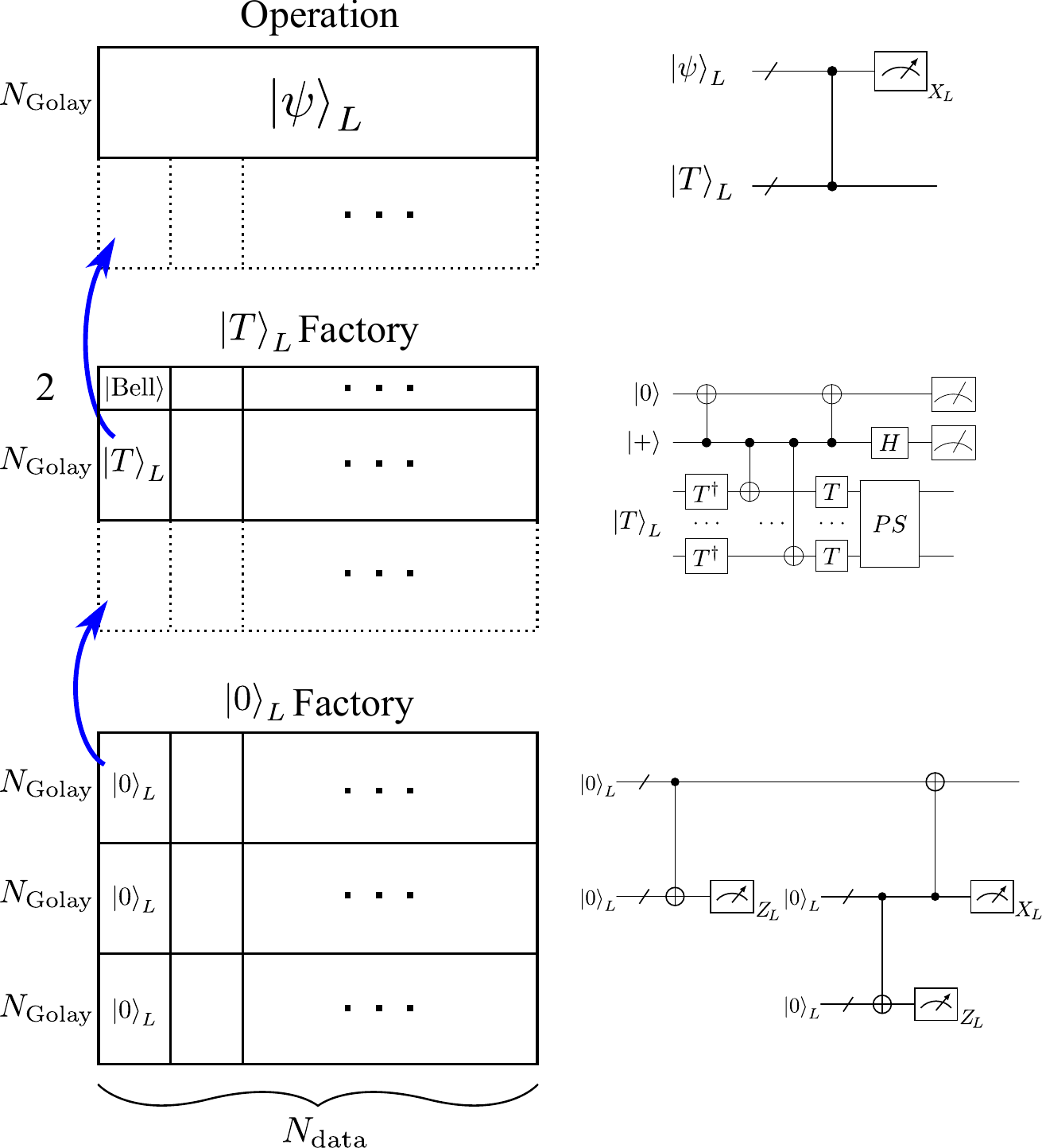}
    \caption{Schematic qubit allocation within the functional three-zone organization
    of MB-FTQC for executing a circuit on $N_{\mathrm{data}}$ logical qubits
    with $\ket{T}$ distillation. Dashed lines indicate regions overlapping with
    preceding areas, which are excluded from the resource estimation.}
    \label{fig:T-factory}
\end{figure}

\begin{samepage}
\noindent
Sharing ancilla factories and optimizing their allocation may further reduce this baseline qubit footprint.
\end{samepage}

\subsection{Error-Budget-Based Quantum-Volume Proxy}
Quantum volume (QV) is a holistic benchmark of a quantum processor's capability to execute wide and deep random circuits~\cite{QV2019}. Higher QV indicates reliable execution of larger and deeper circuits. The square model circuits underlying QV have width $m$ and depth $d=m$ and consist of layers of random two-qubit $\mathrm{SU}(4)$ blocks interleaved with random qubit permutations $\pi$. Because we assume high connectivity, these permutations are treated as cost-free. Following the effective-error approximation of Cross \textit{et al.} and the error-budget estimate used in the STAR architecture~\cite{QV2019,Akahoshi2024}, we estimate the circuit-size threshold at which the accumulated logical-error budget reaches unity, using the resulting continuous value as a coarse proxy for the QV scale.

\subsubsection{Steane code with analog rotation}
The square model circuit contains $m^{2}/2$ two-qubit $\mathrm{SU}(4)$ blocks. Each $\mathrm{SU}(4)$ can be decomposed into three CZs, 15 $R_Z$ rotations, and 15 $H$ gates~\cite{SMB2004} (here we used $U=R_Z R_X R_Z$ Euler decomposition and $R_X = H R_Z H$). We use the reported upper-bound estimates $p_H$ and $p_{\rm CZ}$ for the Clifford gates and the total rotation-error estimate $p_{R_Z}$. Requiring the accumulated logical-error budget to remain below unity gives
\[
\frac{m^{2}}{2}\bigl(15\,p_H + 3\,p_{\rm CZ} + 15\,p_{R_Z}\bigr) < 1 .
\]
Using the estimates at $p=10^{-4}$ (Table~\ref{tab:logical-error-rates}) gives a threshold of $m\simeq56.93$. Rounding this reference value to the nearest integer gives an estimated QV scale of $\log_2\mathrm{QV}\approx57$. The spatial footprint associated with this rounded width follows from $\eta_{\mathrm{Steane}}=35$:
\[
57 \times 35 = 1{,}995 ,
\]
which is within reach of present-day devices.

\subsubsection{Golay code with magic-state distillation}
For the Golay-code implementation, evaluating QV using the standard heavy-output-generation protocol is impractical at the circuit sizes suggested by the estimate below, because identifying the heavy outputs requires classical simulation of the corresponding ideal random circuits. We therefore report a $T$-error-limited upper bound on the square model-circuit size only as a reference scale.

In the QV circuit, each $\mathrm{SU}(4)$ block decomposes into three CNOTs and seven arbitrary single-qubit rotations $U$~\cite{SMB2004}. Using a Clifford+$T$ synthesis, approximating $U$ to accuracy $\delta$ requires $N \approx 3 \log_{2}(1/\delta)$ $T$ gates~\cite{Akahoshi2024}. With error per $T$ gate $p_T=5.360\times10^{-10}$, the error per synthesized $U$ is approximately
\[
p_U \approx N\,p_T \approx 3 p_T \log_{2}(1/\delta).
\]
Equating the accumulation of $T$-gate errors to the target accuracy
($p_U\approx\delta$) gives the self-consistent equation
\[
\delta=3p_T\log_2(1/\delta).
\]
Its solution gives $p_U=\delta\approx3.95439\times10^{-8}$ and
$N\approx\delta/p_T\approx73.78$, or about $74$ $T$ gates per synthesized
rotation.
Since the circuit contains $7m^{2}/2$ such $U$ gates, a coarse error-budget condition
\[
\frac{7 m^{2}}{2}\,p_U < 1
\]
gives $m\approx2{,}688$ as a $T$-error-limited upper bound, rather than a directly evaluated QV value. With $\eta_{\mathrm{Golay}}=117$, the physical-qubit count for this rounded reference width is
\[
2{,}688 \times 117 = 314{,}496 .
\]

As a more concrete reference, a width-100, depth-100 model circuit satisfies this estimated $T$-gate error budget and requires $11{,}700$ physical qubits in our layout. The principal size reference for the Golay track is therefore the characteristic $T$-count scale in the gigaquop regime, rather than a formal QV value.

The Golay implementation has a characteristic $T$-count scale of approximately $1.87\times10^9$ (gigaquop scale), with 117 physical qubits per logical data qubit in our layout. The following representative applications have $T$-gate requirements of the same order as this gigaquop-scale estimate:
\begin{itemize}
  \item \textbf{QPE of FeMoco-54} (Table I, Ref.~\cite{low2025fastquantumsimulationelectronic}): $1.3\times10^{9}$ $T$ gates on $1{,}137$ logical data qubits $\ (\rightarrow 133{,}029$ physical qubits$)$.
  \item \textbf{Factoring RSA-2048} (Table 5, Ref.~\cite{Gidney2025}): $2.8\times10^{9}$ $T$ gates on $1{,}399$ logical data qubits $\ (\rightarrow 163{,}683$ physical qubits$)$.
\end{itemize}
For RSA-2048, Table 5 of Ref.~\cite{Gidney2025} reports an
expected total of $6.5\times10^{9}$ Toffoli gates for a complete
factorization, with 9.2 expected quantum shots. These shots are
separate quantum circuit executions whose measurement outcomes
are combined through classical post-processing. To characterize
the gate-count requirement for a single quantum circuit execution,
we convert the total Toffoli count to an expected per-shot count.
Assuming $1\,\mathrm{Toffoli}=4T$, the corresponding $T$-gate count is $
N_{T,\mathrm{shot}}
\approx \frac{6.5\times10^{9}}{9.2}\times4
\approx 2.8\times10^{9}.
$
This rough estimate yields a \(T\)-gate budget of the same order as the per-shot \(T\)-gate count for RSA-2048. Assessing the feasibility of the full factorization, however, requires a more detailed resource analysis that also accounts for errors from Clifford operations, memory, and other components.

The physical-qubit counts above are simplified estimates for the listed logical widths; architectural optimizations, such as shared ancilla factories, could reduce these footprints. The gate counts illustrate application requirements of the same order as this characteristic $T$-count scale.

\subsection{Comparison with surface-code-based architectures}
\label{subsec:surface-code-resource-comparison}

To assess the resource efficiency of MB-FTQC, we compare Clifford operations and non-Clifford resource-state generation separately with representative surface-code-based architectures, assuming high connectivity for both our proposal and the baselines. We align the physical-error scale and resource definitions as closely as the published data allow, using active spacetime volume that includes the expected cost of failed ancilla-preparation attempts. For Clifford operations, we also compare online critical-path latencies. However, the native gate sets and some details of the noise models differ across architectures.

For Clifford operations, we compare distance-$7$ logical $H$ and CZ/CNOT
between the Golay-code MB-FTQC implementation and two surface-code baselines.
The first retains the same nonlocal transversal-gate footprint but performs
$d=7$ repetitive syndrome-extraction (SE) rounds with minimum-weight perfect
matching (MWPM); the second is transversal STAR with logical matching
(LogMat) decoder, which reduces the extraction to one round using correlated decoding
\cite{ismail2025transversalstararchitecturemegaquopscale}. MB-FTQC implements
CZ, whereas the surface-code schemes implement CNOT.

Tables~\ref{tab:clifford-h-spacetime-comparison}
and~\ref{tab:clifford-cz-spacetime-comparison} summarize the logical error
rate (LER), total active spacetime volume, and critical-path latency.
For this Clifford-gate comparison, the LER is defined as the total probability
$p_{\mathrm{tot}}=\sum_{P\ne I}p_P$ of nonidentity logical Pauli errors
in the effective channel following the ideal gate. The MB-FTQC entries are
upper-bound estimates, as defined in Appendix~\ref{app:simulation-details};
the LogMat and repetitive-SE entries are rough estimates as described in Appendix~\ref{app:clifford-surface-code-comparison}.
The critical-path latencies include decoding time, and the active volumes
include the spacetime cost of holding data qubits for that duration.
For MB-FTQC, the active volumes also include the expected offline factory
cost, including retries after ancilla rejection. The MB-FTQC critical-path
latency estimates assume that verified ancillas are available on demand.

\begin{table*}[t]
\centering
\caption{Logical-$H$ comparison at $d=7$ with aligned physical-error
scales. The comparison assumptions and surface-code LER estimates
are described in Appendix~\ref{app:clifford-surface-code-comparison}.}
\label{tab:clifford-h-spacetime-comparison}
\setlength{\tabcolsep}{6pt}
\renewcommand{\arraystretch}{1.18}
\begin{tabular}{lccc}
\toprule
Architecture / decoder
& $H$ LER
& Active volume (qubit$\cdot$s)
& Critical-path latency (ms) \\
\midrule
MB-FTQC Golay / parity LUT
& $4.331\times10^{-6}$ & 0.4032 & 1.201 \\
Transversal STAR / LogMat
& $4\times10^{-6}$ & 0.1747
& 1.802 \\
Repetitive-SE surface code / MWPM
& $2.5\times10^{-6}$ & 1.2223 & 12.602 \\
\bottomrule
\end{tabular}
\end{table*}

\begin{table*}[t]
\centering
\caption{Logical-CZ/CNOT comparison under the same conditions as
Table~\ref{tab:clifford-h-spacetime-comparison}.  MB-FTQC implements CZ,
whereas the surface-code schemes implement CNOT.  The surface-code LERs are
estimated as described in Appendix~\ref{app:clifford-surface-code-comparison}.}
\label{tab:clifford-cz-spacetime-comparison}
\setlength{\tabcolsep}{6pt}
\renewcommand{\arraystretch}{1.18}
\begin{tabular}{lccc}
\toprule
Architecture / decoder
& CZ/CNOT LER
& Active volume (qubit$\cdot$s)
& Critical-path latency (ms) \\
\midrule
MB-FTQC Golay / parity LUT
& $2.656\times10^{-5}$ & 0.8156 & 1.201 \\
Transversal STAR / LogMat
& $1.3\times10^{-5}$ & 0.3884
& 2.004 \\
Repetitive-SE surface code / MWPM
& $8\times10^{-6}$ & 2.4834 & 12.802 \\
\bottomrule
\end{tabular}
\end{table*}

Compared with the seven-round baseline, the MB-FTQC active volumes for $H$ and
CZ are approximately one third as large. For logical $H$, MB-FTQC and transversal STAR achieve similar estimated
LERs with active volumes of the same order. This comparison uses common connectivity and timing
assumptions, with physical error rates chosen conservatively for MB-FTQC.
Although the noise models are not identical, this result indicates that
MB-FTQC is competitive in spacetime cost with a recent
correlated-decoding-based architecture exploiting nonlocal connectivity.

The critical-path comparison provides a further distinction. With sufficient
factory parallelism to supply ancillas at the required rate, MB-FTQC completes
either logical $H$ or CZ using one online entangling layer, one measurement
layer, and a lightweight LUT lookup. Its critical-path latency is therefore
the shortest among the three schemes considered here: transversal STAR has
approximately $1.5$--$1.7$ times the MB-FTQC latency, whereas the seven-round
repetitive-SE baseline has roughly an order of magnitude greater latency.
A further advantage of the offline-factory organization is its tunable
space--time tradeoff: increasing factory parallelism uses a larger instantaneous qubit
footprint but improves ancilla throughput and reduces online waiting, whereas
reducing the number of factories prioritizes the physical-qubit footprint.

This latency advantage is particularly valuable when wall-clock execution
time is a primary constraint, as in high-connectivity trapped-ion and
neutral-atom platforms with comparatively long two-qubit-gate or measurement
times~\cite{Pino2021,Bluvstein2022}, and in computations with large sequential
circuit depth. Because MB-FTQC requires fewer latency-critical online layers,
the time saved at each logical operation accumulates along the circuit
critical path. The cost model, physical-error-rate assignment, and decoder
assumptions are detailed in
Appendix~\ref{app:clifford-surface-code-comparison}.

\begingroup
\clubpenalty=10000
For non-Clifford resource-state generation, we compare HO-ZLD with the recent
fold-transversal surface-code cultivation protocol, which is designed for
architectures with nonlocal connectivity~\cite{sahay2026foldtransversalcultivation}.
The fold-transversal study benchmarks its protocol against several preceding
cultivation schemes and reports the lowest expected spacetime volume among
the schemes included in that comparison, making it a stringent recent
cultivation-based baseline. At the common nominal physical error rate
$p=10^{-3}$, we compare the same proxy state $\ket{S}_L=S_L\ket{+}_L$ for
both schemes. For each HO-ZLD setting, $r=2$ and $r=3$, we choose the
complementary-gap threshold of the $f=5$ fold-transversal protocol to match
the HO-ZLD proxy-state LER. Table~\ref{tab:nonclifford-spacetime-comparison}
reports the expected volume per accepted proxy resource state in
qubit$\cdot$step. The cost includes failed attempts but excludes subsequent
resource-state consumption and gate teleportation.
\par\endgroup

\begin{table}[t]
\centering
\caption{Expected spacetime volume per accepted $\ket{S}_L$ proxy state at
$p=10^{-3}$, in qubit$\cdot$step.  Each fold-transversal point uses the $f=5$
protocol, with its gap threshold matched to the listed LER, and stops at
$d=11$ immediately after the final post-selection.}
\label{tab:nonclifford-spacetime-comparison}
\footnotesize
\setlength{\tabcolsep}{3pt}
\renewcommand{\arraystretch}{1.18}
\begin{tabular}{lcccc}
\toprule
Setting & LER & HO-ZLD &
\begin{tabular}[c]{@{}c@{}}Fold-\\transversal\end{tabular} &
\begin{tabular}[c]{@{}c@{}}Fold-transversal\\/ HO-ZLD\end{tabular} \\
\midrule
$r=2$ & $1.27\times10^{-7}$ & $10{,}474$ & $18{,}625$ & $1.78$ \\
$r=3$ & $1.45\times10^{-8}$ & $17{,}397$ & $21{,}248$ & $1.22$ \\
\bottomrule
\end{tabular}
\end{table}

Compared with fold-transversal cultivation, HO-ZLD reduces the expected
volume per accepted proxy state by factors of $1.78$ and $1.22$ for
$r=2$ and $r=3$, respectively.
Although a single HO-ZLD preparation attempt has a larger expected volume
than a fold-transversal attempt, its higher acceptance probability reduces
the mean number of attempts enough to yield a lower expected volume per
accepted proxy state.
These results demonstrate that, even after accounting for the cost of
ancilla-preparation retries, HO-ZLD is competitive in spacetime cost
with recent cultivation-based schemes designed for nonlocal connectivity
at matched proxy-state LERs.
The comparison conditions, noise-model differences, and cost calculations
are detailed in
Appendix~\ref{app:nonclifford-fold-transversal-comparison}.

\section{Conclusion}
\label{sec:conclusion}
In this study, we have proposed a novel quantum error correction architecture tailored for highly connected quantum devices with $O(10^3)$ to $O(10^4)$ qubits, anticipated in the near future. Our scheme centers on error-correcting teleportation, enabling logical operations to intrinsically incorporate syndrome extraction and Pauli frame updates. This tight integration allows computation progress and error suppression to occur autonomously and concurrently. As concrete examples, we implemented this framework using self-dual CSS codes, specifically the Steane and Golay codes, to realize a fully measurement-based computational paradigm at the logical layer.

For non-Clifford operations, which are essential for universality, we present two complementary strategies aligned with hardware scale. First, for the Steane code, we adopt a partially fault-tolerant approach based on analog rotations. By carefully designing the ancilla-encoding circuits, we suppress the dominant rotation-induced errors. Including the mean of two RUS attempts, the leading contribution to the logical rotation error is $2p/15$, with $O(p^2)$ corrections including the teleportation contribution. Numerical simulations show $O(p^2)$ scaling for Clifford gates. At $p=10^{-4}$, the estimated total rotation error $p_{R_Z}=2.438\times10^{-5}$ gives approximately $4.10\times10^4$ logical rotations, corresponding to an equivalent $T$-count of $1.89\times10^6$. The error-budget-based QV proxy gives a rounded scale of $\log_2\mathrm{QV}\approx57$, requiring about $1{,}995$ physical qubits in our encoding. This provides a coarse circuit-size reference for the Steane implementation on near-term high-connectivity devices.

Second, for the Golay code, we developed a fully fault-tolerant Clifford+$T$ scheme based on tailored, qubit-efficient physical-level $\ket{T}_L$ distillation, termed \emph{higher-order zero-level distillation}.
Numerical simulations show logical error scaling as $O(p^{4})$ for both Clifford and $T$ gates. At $p=10^{-4}$, the estimated logical $T$-gate error rate is $5.360\times10^{-10}$, corresponding to a characteristic $T$-count scale of $1/p_T\approx1.87\times10^9$ in the gigaquop regime without code concatenation. Within our functional three-zone organization, the resource ratio is $\eta_{\mathrm{Golay}} = 117$ physical qubits per logical qubit, meaning that a device with approximately $1.2 \times 10^4$ qubits can support a width-100, depth-100 model circuit within the estimated $T$-gate error budget. At this scale, $T$-gate errors are not the limiting factor in our estimate. This characteristic $T$-count scale is of the same order as the requirements of applications such as FeMoco QPE and RSA-2048 factoring, whose logical widths correspond to $O(10^5)$ physical qubits in our layout. Our resource estimates are simplified; further optimizations, such as shared or scheduled factory utilization, are expected to reduce the required qubit count.

The comparisons under the same high-connectivity assumption further support
the resource efficiency of the proposed architecture. For Clifford
operations, MB-FTQC substantially reduces both active volume and online
critical-path latency relative to the same-distance
repetitive-syndrome-extraction baseline. It is also comparable in active
volume to transversal STAR under the present conservative normalization,
while providing the
shortest online critical path among the compared schemes when parallel
ancilla factories supply resources at the required rate.
For non-Clifford resource-state generation, HO-ZLD also achieves a lower
expected spacetime cost than the matched fold-transversal counterparts at both
Golay operating points studied, under a common nominal physical-error scale
and matched proxy-state LERs. This establishes an advantage over a recent
high-connectivity cultivation-based baseline within the operating regime used
by the Golay MB-FTQC architecture.

The separation of offline resource preparation from online execution provides
a flexible space--time tradeoff: factory parallelism can be adjusted to
prioritize physical-qubit footprint, sustained throughput, or critical-path
latency. Together with its reduced online circuit depth, this flexibility
makes MB-FTQC particularly attractive for deep computations and for
high-connectivity platforms with long gate and measurement times.

Overall, our architecture offers a promising route to fault-tolerant quantum computation on near-term high-connectivity devices without the overhead of code concatenation.
The underlying framework of logical teleportation with verified ancillas also applies to self-dual CSS codes that support the required transversal operations, including high-rate and qLDPC families.
Implementing the framework with these codes requires code-specific ancilla preparation and verification circuits.
Developing efficient post-selection strategies for these circuits, potentially using decoder confidence, remains an important direction for future work.
Other directions include optimizing factory layouts, accounting for platform-specific noise characteristics, and benchmarking the architecture for applications such as quantum chemistry through end-to-end resource estimates that integrate the full computational error budget with hardware layout and scheduling.

\section*{Acknowledgments}
The authors thank Yugo Takada, Theerapat Tansuwannont, Keita Kanno, and Shoichiro Tsutsui for valuable discussions, and Kentaro Wada for his kind support.
KF is supported by MEXT Quantum Leap Flagship Program (MEXT Q-LEAP) Grant No. JPMXS0120319794, JST COI-NEXT Grant No. JPMJPF2014, JST
Moonshot R\&D Grant No. JPMJMS2061, and  JST CREST JPMJCR24I3.

\bibliography{90_library}

\appendix

\section{Simulation details and effective error models}
\label{app:simulation-details}

\subsection{Circuit-level noise model and decoding}
\label{app:simulation-details-noise-model-and-decoding}

The simulated physical operations are one-qubit Clifford gates, CNOT gates,
initialization to $\ket{0}$, and measurement in the $Z$ basis. We apply the
standard one- and two-qubit depolarizing channels
\begin{equation}
\label{eq:single_depolarizing}
  {\mathcal E}_{\rm single}(\rho)
  = (1-p) \rho
  + \frac{p}{3} \left( X \rho X + Y \rho Y + Z \rho Z \right)
\end{equation}
and
\begin{equation}
\label{eq:double_depolarizing}
  {\mathcal E}_{\rm double}(\rho)
  = (1-p) \rho
  + \frac{p}{15}
  \sum_{E \in \{I,X,Y,Z\}^{\otimes 2}\setminus\{II\}} E\rho E .
\end{equation}
The corresponding channel is applied after each physical one- or two-qubit
gate. A single-qubit depolarizing channel is also inserted after each physical
$\ket{0}$ preparation and immediately before each $Z$-basis measurement to
model initialization and measurement errors, respectively.

As mentioned in Sec.~\ref{sec:introduction}, decoding uses a precomputed
lookup table. For errors of weight below $d/2$, the error configuration is
uniquely determined by its syndrome for the Steane and Golay codes. Enumerating
errors of weights $k=0,1$ for the Steane code and $k=0,1,2,3$ for the Golay
code produces tables with $2^3=8$ and $2^{11}=2048$ entries, respectively.
Because both codes are self-dual, the same table is used for $X$ and $Z$
errors.

\subsection{Logical $\ket{0}$ preparation and effective ancilla model}
\label{subsec:modeling}

\begin{figure}
    \centering
    \includegraphics[width=.95\linewidth]{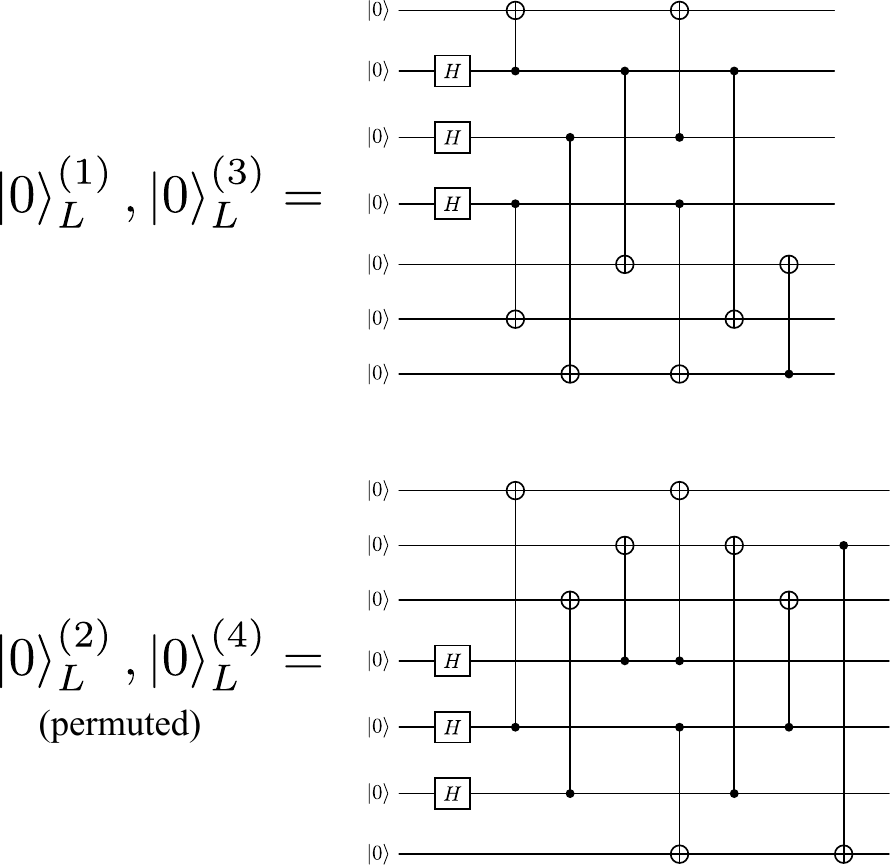}
    \caption{Encoding circuits for the logical $\ket{0}$ state of the Steane
    code, used for the four input blocks of the two-round entanglement
    purification protocol in Sec.~\ref{subsec:method_zero_prep}.}
    \label{fig:steane-encoding-permutation}
\end{figure}

We benchmark the code-specific implementations defined in
Sec.~\ref{subsec:zero-implementation}. The two Steane encoding circuits used
in the four-block verification protocol are shown in
Fig.~\ref{fig:steane-encoding-permutation}.

In the Monte Carlo simulations, post-selection is based on the noisy bitwise
measurement outcomes from code blocks 2, 3, and 4. A sample is discarded if
any stabilizer of these blocks detects an error. To evaluate the accepted
output, we ideally measure its $Z$-type stabilizers and logical $Z$ operator
and infer the corrected logical value. The sample is counted as logically
correct when this value is $+1$.

For the more computationally demanding simulations, we replace explicit
ancilla purification by an effective Pauli error model. The post-selected
output is represented as an ideal $\ket{0}_L$ followed by independent,
identically distributed single-qubit Pauli errors. This form follows because
the encoding-circuit errors are detected by post-selection, while the
remaining first-order faults arise from the final transversal CNOTs in the
purification circuit.

\begin{figure}
    \centering
    \includegraphics[width=0.7\linewidth]{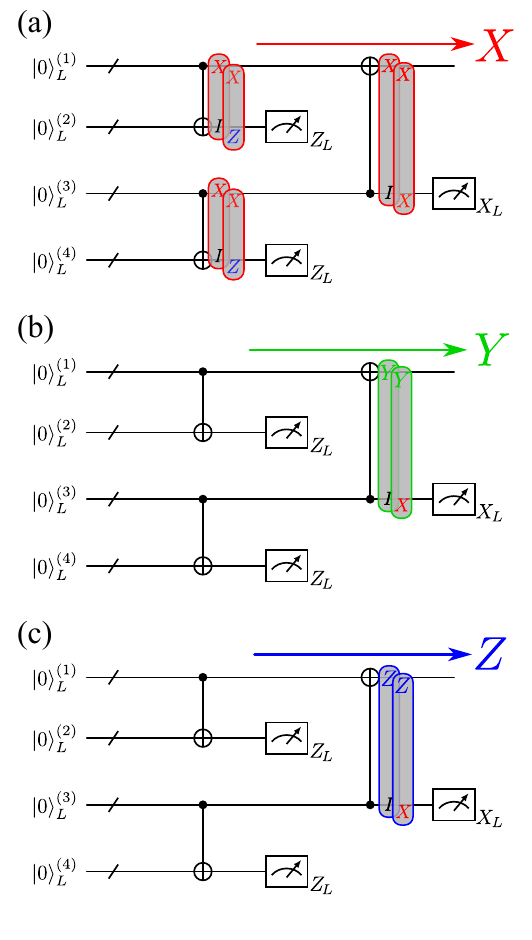}
    \caption{Undetected error patterns after two-round purification of a
    logical $\ket{0}$ state. Six patterns contribute to output $X$ errors,
    while two patterns contribute to each of the output $Y$ and $Z$ errors.}
    \label{fig:purification-error-patterns}
\end{figure}

For each such CNOT, the two-qubit depolarizing channel assigns probability
$p/15$ to every nonidentity two-qubit Pauli. As shown in
Fig.~\ref{fig:purification-error-patterns}, six undetected patterns leave an
$X$ error on the output qubit, while two patterns each leave a $Y$ or $Z$
error. The effective per-qubit probabilities are therefore
\[
p_X=\frac{6}{15}p,\qquad
p_Y=\frac{2}{15}p,\qquad
p_Z=\frac{2}{15}p.
\]

\begin{figure}
    \centering
    \includegraphics[width=.9\linewidth]{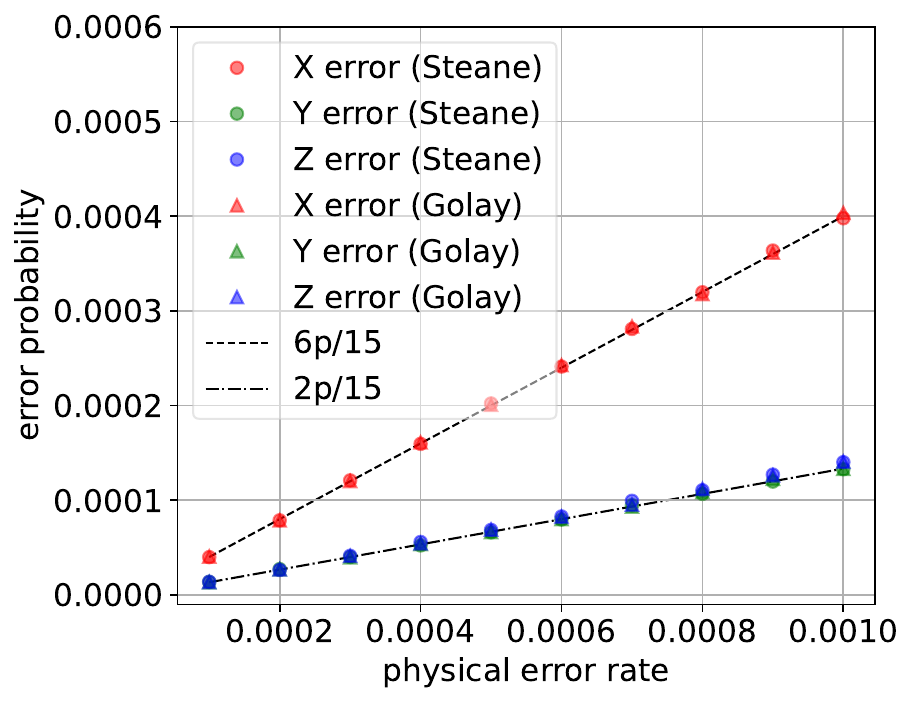}
    \caption{Per-qubit occurrence probabilities of $X$, $Y$, and $Z$ errors
    on post-selected $\ket{0}_L$ states for the Steane (circles) and Golay
    (triangles) codes. The number of samples is $10^7$.}
    \label{fig:xyz-error-dist}
\end{figure}

We verify this model by extracting the per-qubit $X$, $Y$, and $Z$ error
probabilities from post-selected, logically correct samples using their
syndromes. Figure~\ref{fig:xyz-error-dist} shows agreement with the predicted
$6p/15$ and $2p/15$ probabilities for both codes.

\begin{figure}
    \centering
    \includegraphics[width=0.9\linewidth]{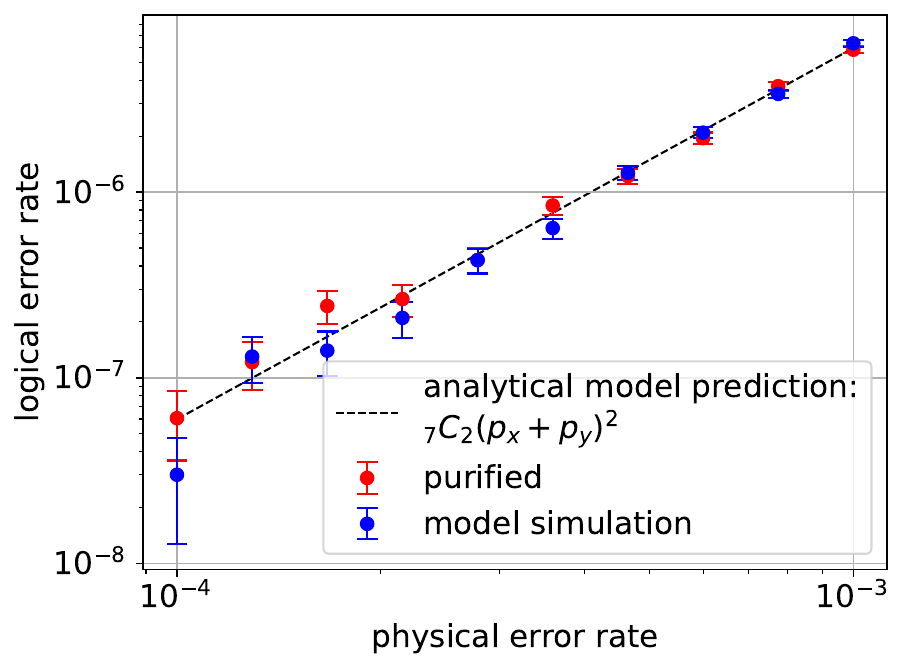}
    \includegraphics[width=0.9\linewidth]{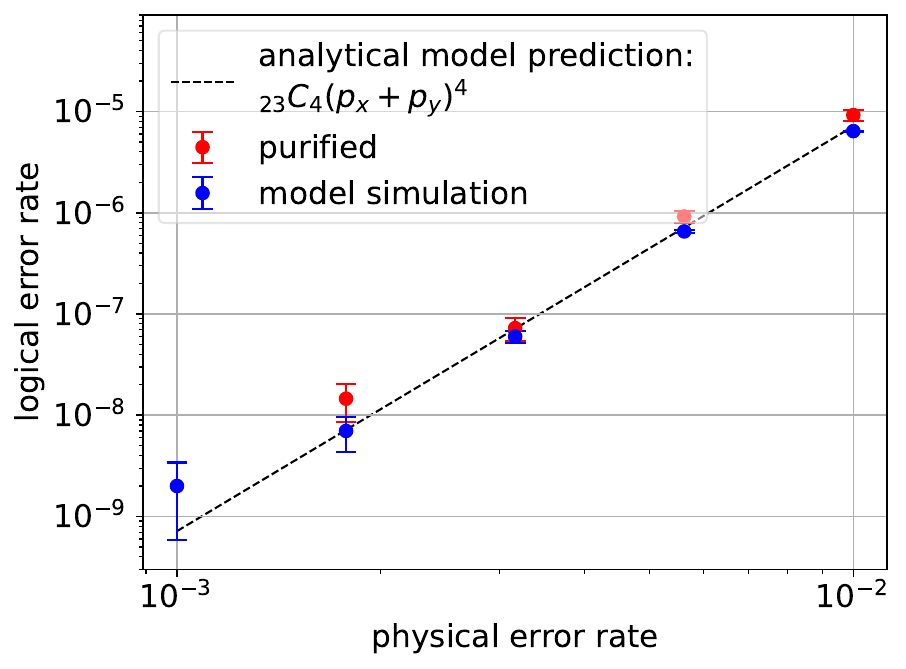}
    \caption{Logical error rates of purified $\ket{0}_L$ states from explicit
    purification, the effective-model simulation, and the leading-order
    analytical prediction. (top) Steane code. (bottom) Golay code. The numbers
    of samples are $10^8$ and $10^9$, respectively.}
    \label{fig:zero-prep-model-comparison}
\end{figure}

At leading order, this transversal model gives
\[
P_L^{\text{Steane}}=\binom{7}{2}(p_X+p_Y)^2,
\qquad
P_L^{\text{Golay}}=\binom{23}{4}(p_X+p_Y)^4.
\]
Figure~\ref{fig:zero-prep-model-comparison} compares these analytical curves
with explicit purification and direct application of the effective channel to
an ideal $\ket{0}_L$. Their agreement supports using the effective model in
the subsequent gate simulations.

\subsection{Repeated-application benchmarks for logical Clifford gates}
\label{appsubsec:repeated-clifford}

We model each logical Clifford gate by its ideal operation followed by an
effective logical Pauli channel. For a given gate, the state-specific error
probabilities below refer to a common channel acting on the ideal output.
For a single logical qubit, we write
\begin{align}
\mathcal{E}(\rho)
&=(1-p_{\mathrm{tot}})\rho
  +\sum_{P\in\{X,Y,Z\}}p_P P\rho P, \nonumber\\
p_{\mathrm{tot}}&=p_X+p_Y+p_Z,
\end{align}
where $P$ denotes a logical Pauli operator. For two logical qubits, the
analogous channel has total nonidentity Pauli error probability
\begin{equation}
p_{\mathrm{tot}}^{(2)}
=\sum_{P\in\{I,X,Y,Z\}^{\otimes2}\setminus\{II\}}p_P.
\end{equation}

For $H$, let $e_0$ and $e_+$ be the probabilities that validation gives an
incorrect logical value when the ideal output is $\ket{0}_L$ and
$\ket{+}_L$, respectively. Then
\begin{equation}
e_0=p_X+p_Y,\qquad e_+=p_Y+p_Z,
\end{equation}
so their sum satisfies
\begin{align}
\overline p_H&=e_0+e_+=p_{\mathrm{tot}}+p_Y,\\
p_{\mathrm{tot}}&\le\overline p_H\le2p_{\mathrm{tot}}.
\end{align}

For CZ, define $\ket{G}_L=\mathrm{CZ}_L\ket{++}_L$. Let $e_{00}$,
$e_{++}$, and $e_G$ denote the probabilities that at least one of the two
independent logical stabilizer generators has an incorrect value for the
corresponding ideal output state. A trial in which both values are
incorrect is counted as one failure. The nonidentity Pauli operators that
leave each state unchanged are
\begin{align*}
\ket{00}_L&:\quad ZI,\ IZ,\ ZZ,\\
\ket{++}_L&:\quad XI,\ IX,\ XX,\\
\ket{G}_L&:\quad XZ,\ ZX,\ YY.
\end{align*}
These three sets are disjoint, so each of the nine listed Pauli errors is
detected for two states. The remaining six errors, $IY$, $YI$, $XY$, $YX$,
$YZ$, and $ZY$, are detected for all three states. Consequently,
\begin{align}
\overline p_{\mathrm{CZ}}
&=\frac{e_{00}+e_{++}+e_G}{2} \nonumber\\
&=p_{\mathrm{tot}}^{(2)}
  +\frac12\bigl(p_{IY}+p_{YI}+p_{XY} \nonumber\\
&\hspace{6em}+p_{YX}+p_{YZ}+p_{ZY}\bigr),\\
p_{\mathrm{tot}}^{(2)}
&\le\overline p_{\mathrm{CZ}}\le\frac32p_{\mathrm{tot}}^{(2)}.
\end{align}

Following Ref.~\cite{Steane2003}, we estimate logical gate errors by repeated
application rather than by testing a single gate on a freshly prepared block.
This retains the physical errors accumulated on the data between successive
logical operations. We use $Q_{\mathrm{max}}=10$ iterations for $H$ and
$Q_{\mathrm{max}}=20$ for CZ, including all initial steps in the analysis.
Let $N(Q)$ count the runs whose first detected logical failure occurs at
step $Q$, out of $N_{\mathrm{shots}}$ runs. We estimate the failure
probability at step $Q$, conditioned on passing all preceding logical
validations, as
\begin{equation}
p(Q)=\frac{N(Q)}{N_{\mathrm{shots}}-\sum_{q=1}^{Q-1}N(q)}.
\label{eq:p_Q}
\end{equation}

\begin{figure}
    \centering
    \includegraphics[width=0.9\linewidth]{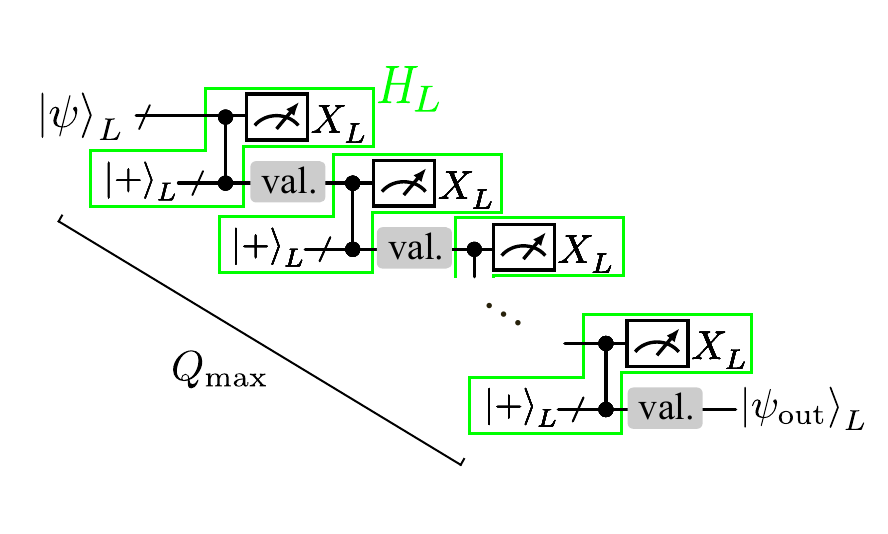}
    \caption{Repeated-evaluation circuit for logical $H$ gates. ``val.''
    denotes validation by Pauli-product measurements. The sequence is
    repeated up to $Q_{\mathrm{max}}$ times.}
    \label{fig:logical-h-eval}
\end{figure}

For the logical $H$ benchmark, one round consists of an LOBT operation
followed by ideal validation, as shown in Fig.~\ref{fig:logical-h-eval}. The
input is $\ket{\psi}_L=\ket{+}_L$. At each validation, we measure the
appropriate stabilizers and logical operator, infer the corrected logical
value, and compare it with the expected result. Odd and even steps ideally
produce $\ket{0}_L$ and $\ket{+}_L$, respectively, so logical $Z$ and $X$ are
measured in alternation. The LOBT outcome $m\in\{0,1\}$ is retained in the
Pauli frame when evaluating the output.

We take the maximum of $p(Q)$ separately over the odd and even steps:
\begin{align}
\widehat e_0
&=\max_{\substack{1\le Q\le Q_{\mathrm{max}}\\Q\ \mathrm{odd}}}p(Q),
\nonumber\\
\widehat e_+
&=\max_{\substack{1\le Q\le Q_{\mathrm{max}}\\Q\ \mathrm{even}}}p(Q).
\end{align}

\begin{figure*}[thbp]
    \centering
    \includegraphics[width=.9\textwidth]{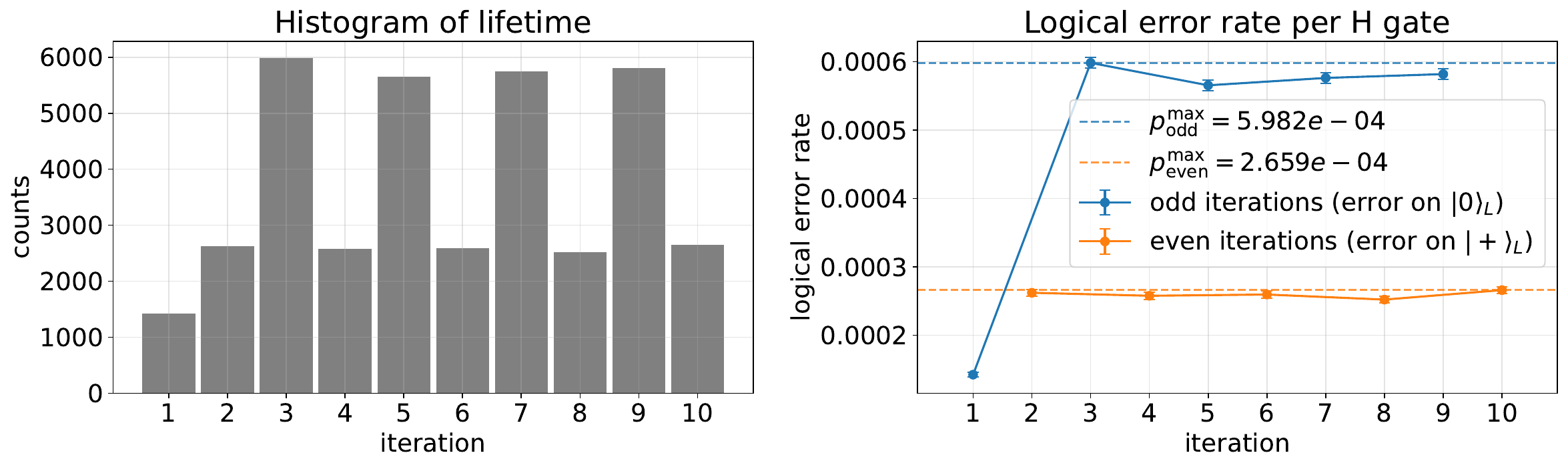}
    \includegraphics[width=.9\textwidth]{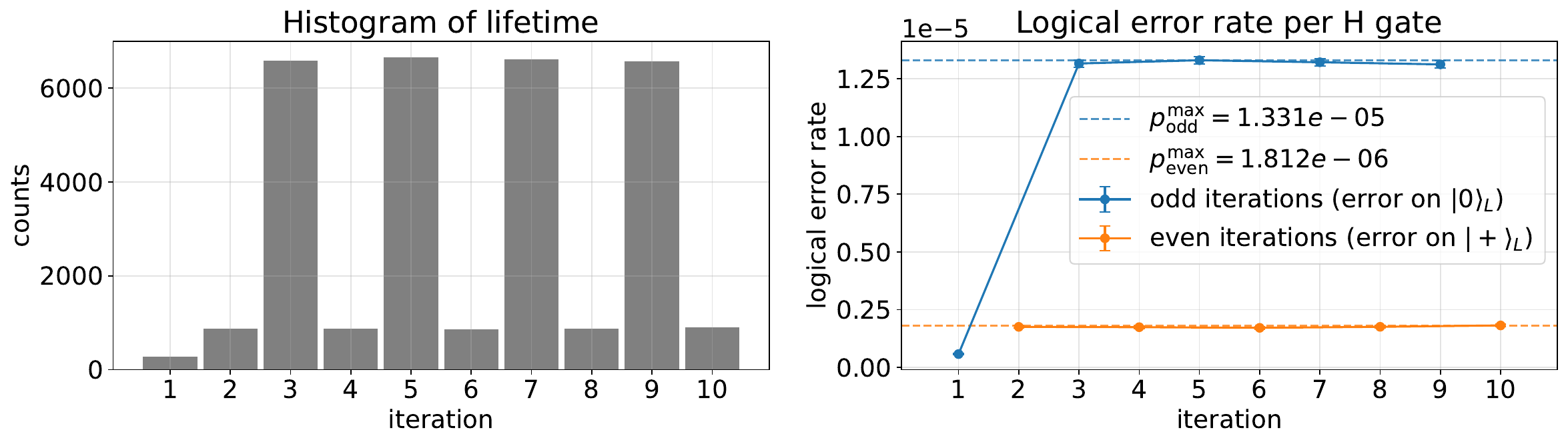}
    \caption{Logical $H$ gate error-rate estimation for the Steane code (top)
    and Golay code (bottom) at $p=10^{-3}$. The numbers of samples are $10^7$
    and $5\times10^8$, respectively.}
    \label{fig:logical-h-hist}
\end{figure*}

Figure~\ref{fig:logical-h-hist} shows the initial transient and the subsequent
even--odd oscillation of $p(Q)$. Errors in the measured LOBT block can change
the inferred outcome $m$ and hence the tracked $X^m$ byproduct. At even steps,
the output is $\ket{+}_L$, on which this byproduct acts trivially, producing
the lower branch of the oscillation.

\begin{figure*}[thbp]
    \centering
    \includegraphics[width=.9\textwidth]{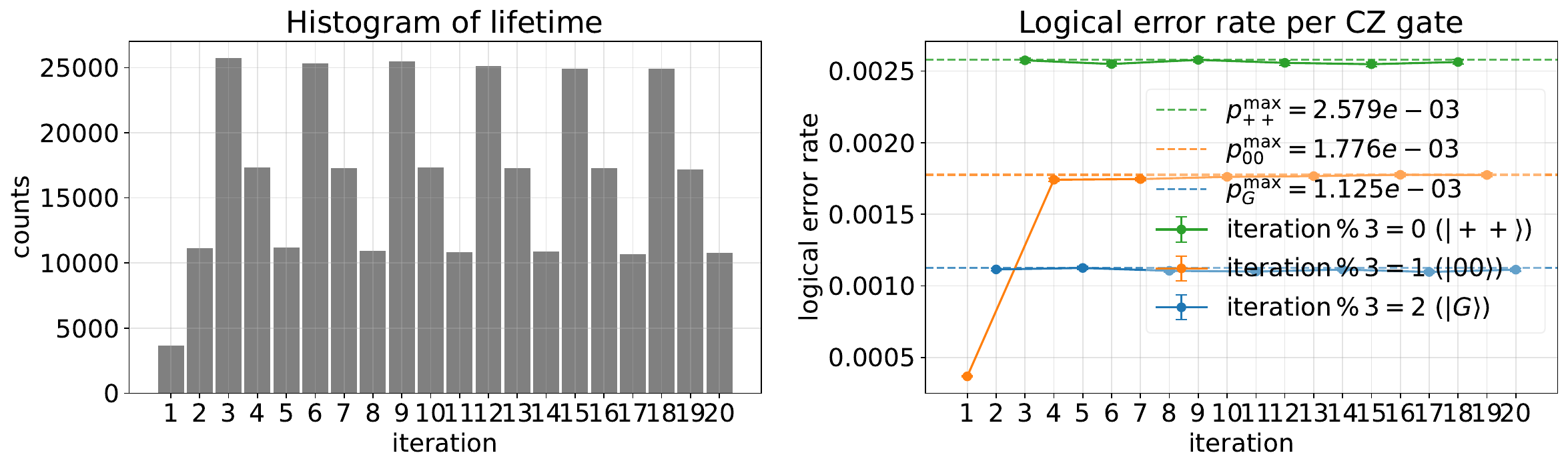}
    \includegraphics[width=.9\textwidth]{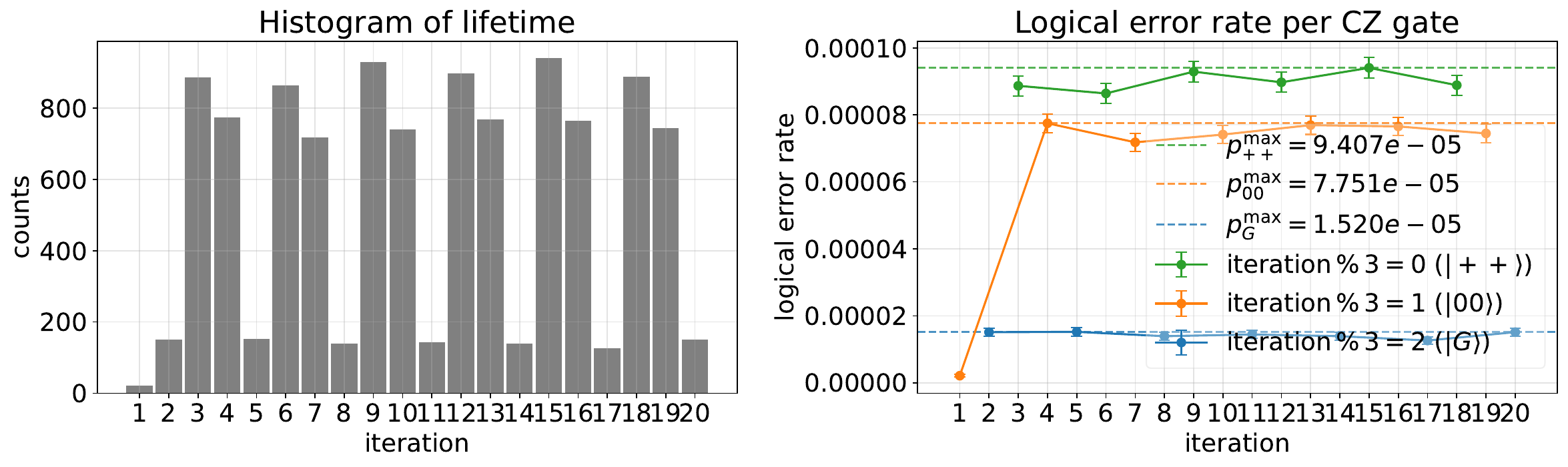}
    \caption{Logical CZ gate error-rate estimation for the Steane code (top)
    and Golay code (bottom) at $p=10^{-3}$. The number of samples is $10^7$
    for each.}
    \label{fig:logical-cz-hist}
\end{figure*}

For the logical CZ benchmark, the initial state is $\ket{++}_L$, and each
round applies $U=\mathrm{CZ}(H\otimes H)$. Since
$U^3=\mathrm{SWAP}$, the two commuting stabilizers evolve with period three:
$\{XI,IX\}\to\{ZI,IZ\}\to\{XZ,ZX\}\to\{IX,XI\}\to\cdots$. We validate
both stabilizers after each round, including the Pauli-frame byproducts, and
count a failure when at least one corrected logical value is incorrect.
The ideal output states are $\ket{00}_L$, $\ket{G}_L$, and $\ket{++}_L$
for $Q\equiv1,2,0\pmod3$, respectively. Using Eq.~\eqref{eq:p_Q}, we take
the maximum within each of these three phases:
\begin{align}
\widehat e_{00}
&=\max_{\substack{1\le Q\le Q_{\mathrm{max}}\\Q\equiv1\pmod3}}p(Q),
\nonumber\\
\widehat e_G
&=\max_{\substack{1\le Q\le Q_{\mathrm{max}}\\Q\equiv2\pmod3}}p(Q),
\nonumber\\
\widehat e_{++}
&=\max_{\substack{1\le Q\le Q_{\mathrm{max}}\\Q\equiv0\pmod3}}p(Q).
\end{align}
The period-three oscillation with the initial transient is shown in
Fig.~\ref{fig:logical-cz-hist}.

We report the numerical estimates
\begin{align}
\widehat{\overline p}_H&=\widehat e_0+\widehat e_+, \nonumber\\
\widehat{\overline p}_{\mathrm{CZ}}
&=\frac{\widehat e_{00}+\widehat e_{++}+\widehat e_G}{2}
\label{eq:clifford-pauli-upper-bounds}
\end{align}
in Fig.~\ref{fig:Clifford}. Interpreting these as upper-bound estimates
assumes that, for each gate, the phase-resolved maxima estimate the
state-specific error probabilities of a common effective logical Pauli
channel. The bounds derived above apply to that channel.

The benchmarks use the effective ancilla model of
Sec.~\ref{subsec:modeling}. For the Steane $H$ gate, we additionally simulate
explicitly purified ancillas; the agreement shown in Fig.~\ref{fig:Clifford}
validates the model for logical gate evaluation.

\subsection{Teleportation contributions for non-Clifford gates}

For both codes, the teleportation benchmarks use the same sum of the two
state-specific error estimates as the $H$-gate benchmark. We denote the
resulting upper-bound estimate per teleportation attempt by
$p_{\mathrm{tele}}$.

\begin{figure}
    \centering
    \includegraphics[width=0.9\linewidth]{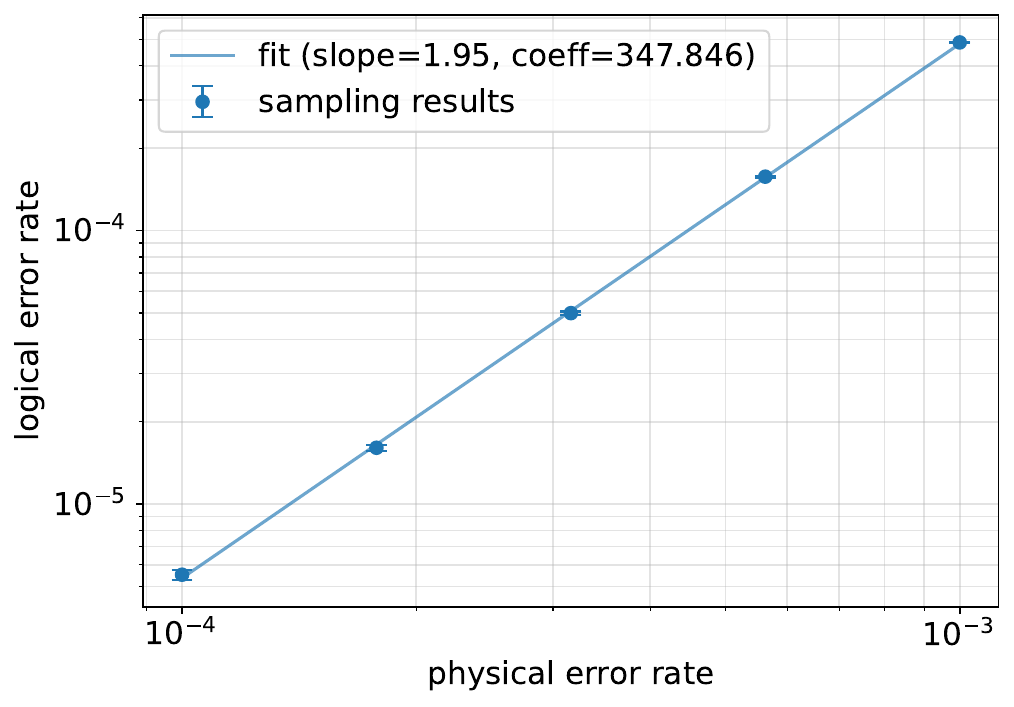}
    \caption{Estimated upper bound on the gate-teleportation error contribution to
    the Steane-code analog $R_Z(\theta)_L$ rotation. The number of samples is
    $10^8$.}
    \label{fig:steane-rz-tele}
\end{figure}

For the Steane analog rotation, we evaluate the teleportation contribution by
repeated LOBT applications using an ideal $\ket{+}_L$ ancilla followed by
independent physical $X$, $Y$, and $Z$ errors with probabilities $2p/15$,
$2p/15$, and $10p/15$, respectively, as obtained from error-propagation
analysis. The logical state-preparation contribution is evaluated
separately as $p_{\mathrm{rot}}$ and is excluded from this benchmark. The
resulting estimate $p_{\mathrm{tele}}$ scales as $O(p^2)$, as shown in
Fig.~\ref{fig:steane-rz-tele}.
Table~\ref{tab:steane-rz-error-breakdown} gives the contributions to
$p_{R_Z}\approx2(p_{\mathrm{rot}}+p_{\mathrm{tele}})$ at the physical error
rates used in Tables~\ref{tab:logical-error-rates-1e-3}
and~\ref{tab:logical-error-rates}. The factor of two accounts for the mean
number of RUS attempts per logical rotation.

\begin{table}[tb]
    \centering
    \caption{Breakdown of the estimated logical $R_Z$-rotation error rate
    for the Steane code. Here $p_{\mathrm{rot}}$ is the state-preparation
    contribution and $p_{\mathrm{tele}}$ is an upper-bound estimate of the
    teleportation error probability per attempt.}
    \label{tab:steane-rz-error-breakdown}
    \small
    \setlength{\tabcolsep}{3pt}
    \renewcommand{\arraystretch}{1.2}
    \begin{tabular}{@{}cccc@{}}
        \toprule
        $p$ & $p_{\mathrm{rot}}$ & $p_{\mathrm{tele}}$ & $p_{R_Z}$ \\
        \midrule
        $10^{-3}$ & $1.051\times10^{-4}$ & $4.871\times10^{-4}$
        & $1.184\times10^{-3}$ \\
        $10^{-4}$ & $6.670\times10^{-6}$ & $5.520\times10^{-6}$
        & $2.438\times10^{-5}$ \\
        \bottomrule
    \end{tabular}
\end{table}

\begin{figure}
    \centering
    \includegraphics[width=0.9\linewidth]{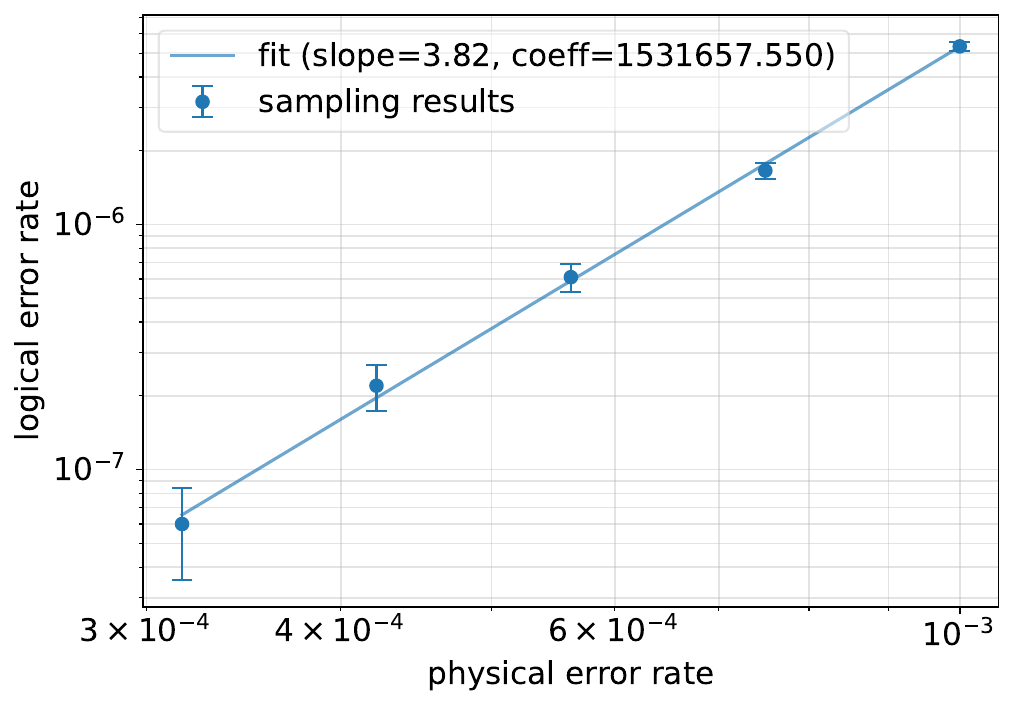}
    \caption{Estimated upper bound on the gate-teleportation error contribution to
    the Golay-code $T$ gate. The number of samples is $10^8$.}
    \label{fig:golay-t-tele}
\end{figure}

For the Golay HO-ZLD simulation, the verified $\ket{0}_L$ inputs to Steane's
gadget are represented using the effective model of
Sec.~\ref{subsec:modeling}. The transversal gate-teleportation contribution is
simulated independently of the $\ket{S}_L$ proxy-state preparation; the
upper-bound estimate is shown in Fig.~\ref{fig:golay-t-tele}.
Table~\ref{tab:golay-t-error-breakdown} gives the contributions to
$p_T\approx2p_{\ket{S}}+p_{\mathrm{tele}}$ for the parameter choices used in
Tables~\ref{tab:logical-error-rates-1e-3} and~\ref{tab:logical-error-rates}.
The factor of two applies only to the proxy-state preparation contribution.

\begin{table}[tb]
    \centering
    \caption{Breakdown of the estimated logical $T$-gate error rate for the
    Golay code. Here $r$ is the number of verification repetitions and
    $p_{\ket{S}}$ is the logical error rate of the accepted $\ket{S}_L$
    proxy state, while $p_{\mathrm{tele}}$ is an upper-bound estimate of the
    teleportation error probability. The entries at $p=10^{-4}$ are
    extrapolated from fits.}
    \label{tab:golay-t-error-breakdown}
    \small
    \setlength{\tabcolsep}{3pt}
    \renewcommand{\arraystretch}{1.2}
    \begin{tabular}{@{}ccccc@{}}
        \toprule
        $p$ & $r$ & $p_{\ket{S}}$ & $p_{\mathrm{tele}}$ & $p_T$ \\
        \midrule
        $10^{-3}$ & $2$ & $1.272\times10^{-7}$ & $5.330\times10^{-6}$
        & $5.584\times10^{-6}$ \\
        $10^{-4}$ & $3$ & $1.450\times10^{-12}$ & $5.331\times10^{-10}$
        & $5.360\times10^{-10}$ \\
        \bottomrule
    \end{tabular}
\end{table}

\section{Estimate of memory errors}
\label{app:memory-errors}

For trapped-ion and neutral-atom devices, memory errors on the timescale
of a two-qubit gate are expected to be much smaller than two-qubit gate
errors. As a rough estimate, we take
$p_{\mathrm{idle}} \sim t_{2Q}/T_{\mathrm{coh}}$, where
$T_{\mathrm{coh}}$ is the measured coherence time and $t_{2Q}$ is the
two-qubit gate duration. Representative experimental values are

\begin{center}
\begin{tabular}{lcccc}
\hline
Platform & $T_{\mathrm{coh}}$ & $t_{2Q}$ & $p_{2Q}$
& $p_{\mathrm{idle}}/p_{2Q}$ \\
\hline
Trapped ion~\cite{Pino2021}
& $2.0\,\mathrm{s}$
& $25\,\mu\mathrm{s}$
& $7.9\times10^{-3}$
& $1.6\times10^{-3}$ \\
Neutral atom~\cite{Bluvstein2022}
& $4\,\mathrm{ms}$
& $0.4\,\mu\mathrm{s}$
& $2.6\times10^{-2}$
& $3.8\times10^{-3}$ \\
\hline
\end{tabular}
\end{center}

Here, the trapped-ion coherence time is a spin-echo $T_2$, while the
neutral-atom value is the inhomogeneous dephasing time $T_2^*$.
In both representative devices, $p_{\mathrm{idle}}/p_{2Q}<10^{-2}$.
Thus, for idle periods comparable to a two-qubit gate duration, memory
errors are substantially smaller than two-qubit gate errors, supporting
their neglect in our circuit-level simulations.

\section{Sensitivity to memory errors}
\label{app:memory-error-sensitivity}

To directly assess the sensitivity of our results to memory errors, we
repeat the simulations of the Steane-code encoding circuit while
applying a memory error at every idle time step. We consider
$p_{\mathrm{idle}}=p/10$ and $p_{\mathrm{idle}}=p$, where $p$ is the
physical gate error rate used in the original simulations, and compare
them with the original case without memory errors.

\begin{figure}[htbp]
    \centering
    \includegraphics[width=0.9\linewidth]{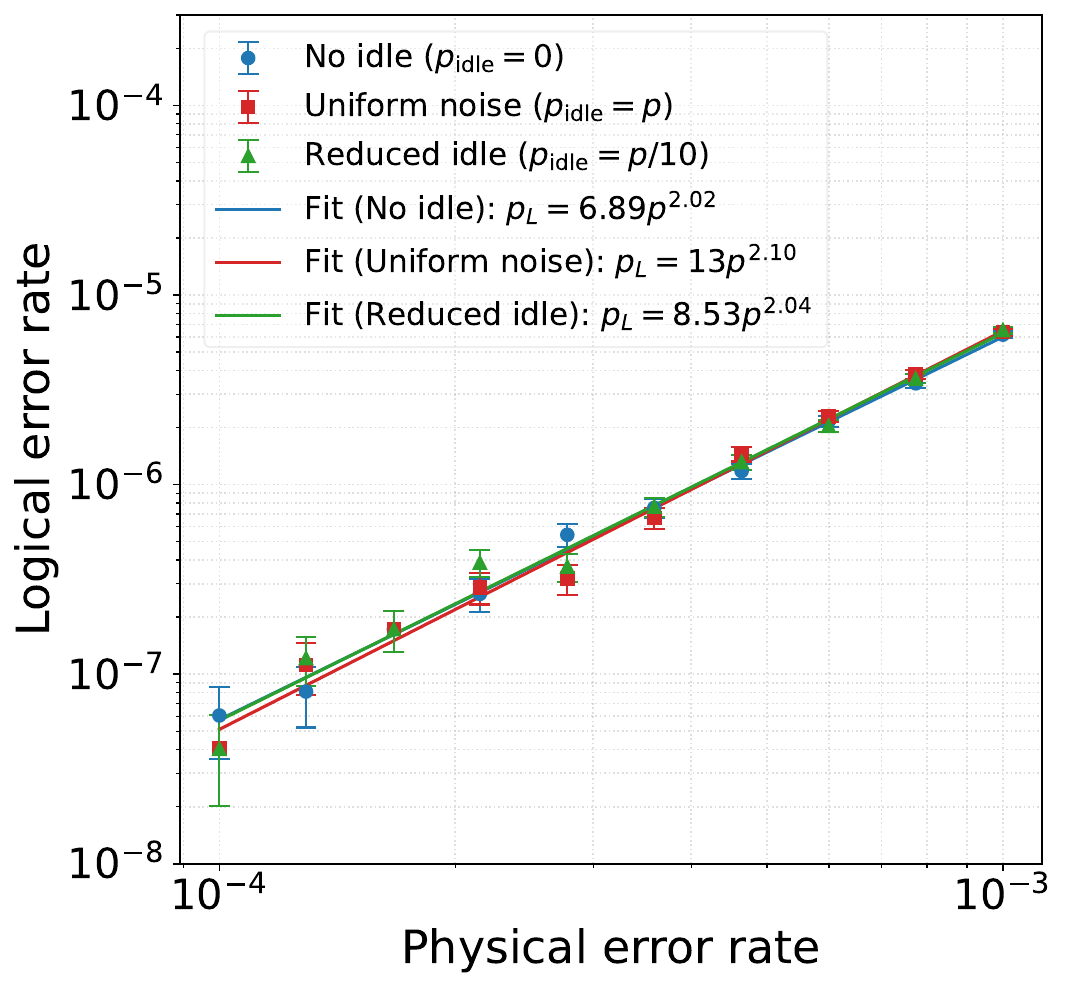}\\[0.5cm]
    \includegraphics[width=0.9\linewidth]{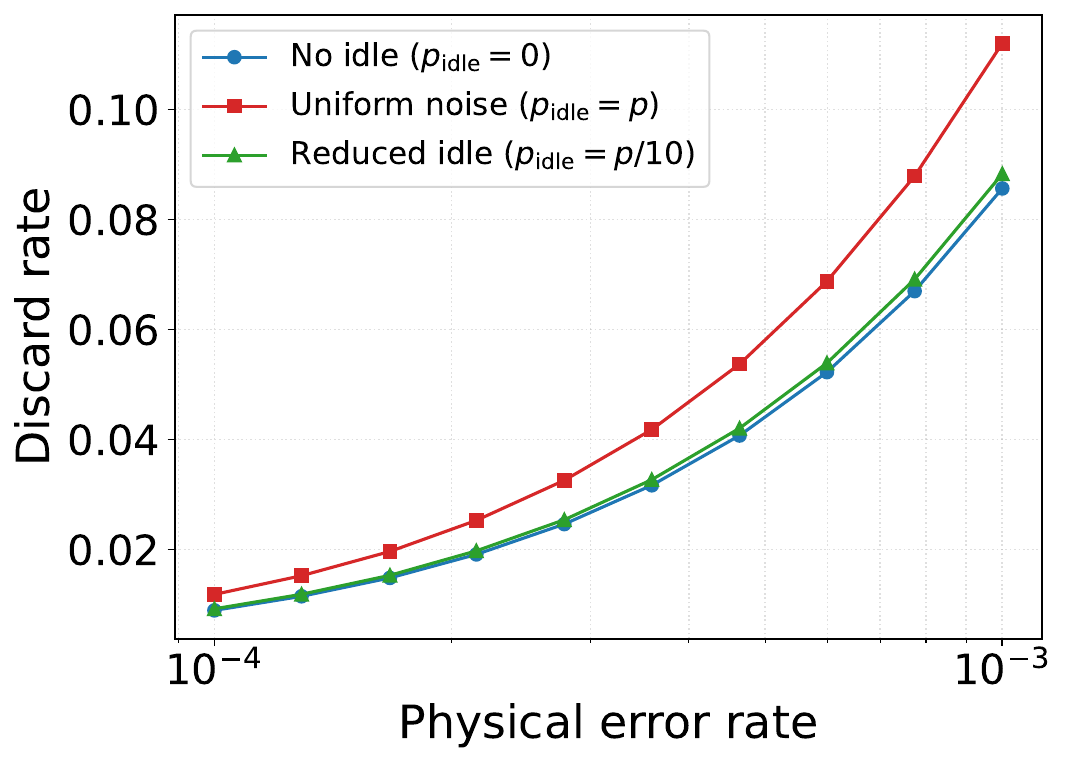}
    \caption{Sensitivity of the Steane-code encoding circuit to memory errors, comparing $p_{\mathrm{idle}}=0$, $p/10$, and $p$. The number of samples is $10^8$. (top) Logical error rate. (bottom) Discard rate. }
    \label{fig:memory-sensitivity}
\end{figure}

Figure~\ref{fig:memory-sensitivity} shows the logical error rate (top) and discard rate (bottom). Including memory errors produces no appreciable change in the logical error rate within statistical uncertainty, even for the deliberately pessimistic choice $p_{\mathrm{idle}}=p$. The main effect is instead an increase in the
discard rate. This effect remains modest for $p_{\mathrm{idle}}=p/10$.
Even in the worst-case setting $p_{\mathrm{idle}}=p$ and $p=10^{-3}$, the post-selection performance changes only
moderately, corresponding to approximately a $2.6\%$ reduction in the success probability.

These results indicate that memory errors primarily affect the
post-selection overhead rather than the logical fidelity of the
accepted states. Since experimentally relevant memory errors are
expected to be much smaller than two-qubit gate errors, as discussed
in Appendix~\ref{app:memory-errors}, neglecting them in the main
simulations does not affect our qualitative conclusions.

\section{Error mechanism and the $p/15$ bound for analog-rotation ancillas on the Steane code}\label{appsec:analog-rot}

Here we describe how a prepared state $\ket{+_\theta}_L$ exhibits a logical error rate of $p/15$ to first order in $p$ (see Fig.~\ref{fig:analog_rotation_circuit}). Throughout, we assume that the only first-order noise acts during $R_{ZZ}(\theta)$ and is modeled as a two-qubit depolarizing channel with total strength $p$, so that each non-identity two-qubit Pauli occurs with probability $p/15$. Errors from all other sources are assumed to be detected by syndrome measurements.

\begin{enumerate}
\item After the first two CNOTs ($\mathrm{CNOT}_{2, 0}$ and $\mathrm{CNOT}_{3, 1}$), the four involved qubits (0--3) host a state stabilized by ${XXXX, ZIZI, IZIZ}$, which is equivalent to the logical $\ket{+}$ of the $[[4,1,2]]$ error-detecting code.
\item $R_{ZZ}(\theta)$ acts on qubits 0 and 1. Among the 15 Pauli errors of the two-qubit depolarizing channel, only $ZZ$ (i.e., $ZZII$ on qubits 0--3) \emph{commutes} with those stabilizers and induces a logical $Z$; the other 14 \emph{anticommute} with at least one of ${XXXX, ZIZI, IZIZ}$, flipping a syndrome.
\item The remaining six CNOTs map these early stabilizers into (a subset of) the stabilizer group of the Steane code. Conjugation by the same circuit preserves (anti)commutation, so every non-$ZZ$ fault continues to flip some Steane stabilizer.
\item Steane's gadget measures three $X$- and three $Z$-type syndromes; any flipped stabilizer triggers rejection. The sole first-order survivor is the $ZZ$ fault on $R_{ZZ}$, which is indistinguishable from a logical $R_Z$ rotation and hence unflagged by syndromes.
\end{enumerate}
Consequently, accepted ancillas have a residual logical $Z$ error with probability $p/15$ at leading order.

\section{Latency of parallel repeat-until-success rotations}
\label{app:rus-latency}

This appendix analyzes how the stochastic completion times of parallel
repeat-until-success (RUS) rotations affect the latency of a multi-qubit
rotation layer. Each logical $R_Z(\theta)_L$ rotation is implemented by
RUS teleportation, and here we quantify the additional circuit latency
caused by the random number of RUS rounds. We assume that the required
resource states are supplied when needed; their preparation and
post-selection retry costs are accounted for separately in
Sec.~\ref{subsec:postselection-overhead}.

For a single logical rotation, let $K$ be the number of RUS rounds until
success. Each round succeeds with probability $1/2$, and hence
\[
  \Pr[K=k]=2^{-k},
  \qquad k=1,2,\ldots,
  \qquad
  \mathbb{E}[K]=2.
\]
This is the sense in which a logical rotation has an amortized latency
of two RUS rounds.

To quantify the possible dragging effect in a multi-qubit circuit, we
consider a conservative synchronized-barrier model. In one rotation
layer, $N$ logical rotations are performed in parallel, and the next
entangling layer starts only after all $N$ rotations have succeeded.
Assuming independent RUS outcomes, the rotation-layer latency is
\[
  X_N=\max_{1\leq i\leq N}K_i,
\]
where the $K_i$ are independent geometric random variables with success
probability $1/2$. For an integer $k\geq0$,
\[
  \Pr[X_N>k]
  =
  1-\left(1-2^{-k}\right)^N
  \leq N2^{-k},
\]
and therefore
\[
  \mathbb{E}[X_N]
  =
  \sum_{k=0}^{\infty}
  \left[
    1-\left(1-2^{-k}\right)^N
  \right].
\]
Although the geometric distribution has unbounded support and thus
does not admit a finite deterministic worst-case delay, its tail is
exponentially suppressed. For large $N$, the expected synchronized
latency scales as
\[
  \mathbb{E}[X_N]
  \simeq
  \log_2 N+\frac{\gamma}{\ln 2}+\frac{1}{2},
\]
up to a small bounded lattice correction. Thus, the expected latency
penalty from global synchronization grows only logarithmically with the
number of simultaneous rotations.

For example, for $N=100$,  $\mathbb{E}[X_{100}]\simeq7.98$.
This corresponds to a deliberately dense rotation layer in which all
100 logical qubits undergo an RUS rotation simultaneously. Relative to
the mean latency $\mathbb{E}[K]=2$ of an isolated rotation, the expected
rotation-layer latency is increased by a factor of approximately
$7.98 / 2\simeq4.0$ under the synchronized-barrier assumption.

The global barrier is conservative. With asynchronous dataflow
scheduling, an entangling operation can be executed once its operand
qubits are ready, while operations that do not depend on a delayed
qubit need not wait. The impact on the total circuit runtime therefore
depends on the circuit dependency graph and compiler schedule, and can
be smaller than the global-barrier estimate.

Under the fixed factory provisioning and qubit allocation assumed in
our resource estimate, random RUS latency increases wall-clock time and
spacetime volume but does not change the physical-qubit footprint.
Synchronization also does not increase the mean number of RUS rounds
performed on each logical qubit: the expected total number of
rotation-teleportation attempts across $N$ rotations remains $2N$.
Consequently, the intrinsic logical-error estimate for each RUS
rotation, which already includes $\mathbb{E}[K]=2$, is unchanged by the
synchronization model. Errors accumulated by qubits while waiting
constitute a separate memory-error contribution. These errors are
expected to be small in the intended trapped-ion and neutral-atom
regime, as quantified in Appendices~\ref{app:memory-errors}
and~\ref{app:memory-error-sensitivity}.

\section{Clifford-gate spacetime-cost comparison with surface-code-based architectures}
\label{app:clifford-surface-code-comparison}

This appendix details the assumptions and cost derivations underlying the
Clifford-gate comparisons in
Tables~\ref{tab:clifford-h-spacetime-comparison}
and~\ref{tab:clifford-cz-spacetime-comparison} of
Sec.~\ref{subsec:surface-code-resource-comparison}.

\subsection{Common assumptions}

We evaluate the three schemes at the common scalar comparison point
\begin{equation}
 p_{\rm MB}=p_{\rm ph}=7.3\times10^{-4},
 \label{eq:mb-transversal-star-error-normalization}
\end{equation}
using the hardware clock adopted for transversal STAR.  In that architecture,
$p_{\rm ph}$ is the combined error probability of a CZ implementation,
including transport and loss contributions
\cite{ismail2025transversalstararchitecturemegaquopscale},
\begin{equation}
 p_{\rm ph}
 =p_{\rm CZ}+2p_{\rm CZ,loss}+p_{\rm move}+p_{\rm move,loss}.
 \label{eq:transversal-star-combined-physical-error}
\end{equation}
At the tenfold-improved operating point used in the transversal STAR study,
the initialization and one-qubit-gate error rates are $3\times10^{-4}$ and
the readout error rate is $4\times10^{-4}$, all below
$p_{\rm ph}=7.3\times10^{-4}$.  By contrast, the MB-FTQC circuit-level
depolarizing model assigns the full $p_{\rm MB}$ to every preparation,
one-qubit gate, two-qubit gate, and measurement location.  The equality in
Eq.~\eqref{eq:mb-transversal-star-error-normalization} therefore assigns the
combined CZ-level error rate also to the non-CZ locations of MB-FTQC and is,
in this sense, conservative for MB-FTQC and favorable to transversal STAR.
Conversely, the transversal STAR model includes spectator errors that are
absent from the MB-FTQC model; the equality is thus a useful scalar
normalization, not an exact equivalence or stochastic ordering of the two
noise channels.  For all three schemes, we adopt the hardware timing
parameters used in the transversal STAR study,
\begin{equation}
 t_{\rm CZ}=0.2\,\mathrm{ms},
 \qquad
 t_M=1\,\mathrm{ms}.
\end{equation}
For the two surface-code baselines, one SE round comprises four CZ sublayers
and one measurement layer, giving
\begin{equation}
 \tau_{\rm SE}=4t_{\rm CZ}+t_M=1.8\,\mathrm{ms}.
 \label{eq:clifford-se-time}
\end{equation}
Physical one-qubit-gate times are neglected, and the entangling layer of a
CNOT is counted separately from SE.

The reported resource totals include decoder computation, although its
contribution is small compared with the quantum-operation times under these
assumptions.  We define
\begin{align}
 V_{\rm circ}&\equiv\sum_j N_jt_j,
 &V_{\rm total}=V_{\rm circ}+N_{\rm hold}t_{\rm dec},
 \nonumber\\
 T_{\rm total}&=T_{\rm circ}+t_{\rm dec},
 \label{eq:active-spacetime-volume}
\end{align}
where $N_j$ qubits are active for time $t_j$ and $N_{\rm hold}$ data qubits
retain the output during decoding.  For MB-FTQC, $V_{\rm circ}$ also includes
the expected offline factory work.  For MB-FTQC, $N_{\rm hold}=23$ for $H$
and 46 for CZ.  For the surface-code schemes, $N_{\rm hold}=49$ for $H$ and
98 for CNOT.
We use per-gate decoding times of $1\,\mu\mathrm{s}$ for the MB parity LUT,
$2.1\,\mu\mathrm{s}$ for transversal STAR/LogMat $H$,
$4.3\,\mu\mathrm{s}$ for transversal STAR/LogMat CNOT, and
$2\,\mu\mathrm{s}$ for the repetitive-SE MWPM baseline; no tail-latency
multiplier is applied.  The sources and assumptions behind these decoding-time
estimates are described below.  Critical-path latency is measured from the start of
the online gadget until its decoder result is available.
The decoder times represent computational runtimes and exclude communication
and classical-control latency.

\subsection{Cost derivation}

The MB-FTQC active volume is the sum of the expected verified-ancilla
preparation cost and the gate-teleportation cost.  Let
$N_{\rm Golay}=23$.  Each verified-ancilla factory uses four Golay blocks,
with seven CNOT layers for encoding and two further layers for purification,
for a total of nine CNOT layers.  Ancilla post-selection is stochastic, so the
factory contribution per accepted resource is divided by the acceptance
probability.  Exponential interpolation of the simulated values
$p_s(10^{-3})=0.6068$ and $p_s(10^{-4})=0.9512$ gives
$p_s(7.3\times10^{-4})\simeq0.6944$.
The resulting circuit volumes are
\begin{align}
 V_{\rm MB,H}^{\rm circ}
 &=\frac{4N_{\rm Golay}(9t_{\rm CZ}+t_M)}{p_s}
 \nonumber\\
 &\quad+N_{\rm Golay}(2t_{\rm CZ}+t_M)
 \nonumber\\
 &\simeq0.4032\ \mathrm{qubit}\mathbin{\cdot}\mathrm{s},
 \label{eq:mb-h-active-volume}\\
 V_{\rm MB,CZ}^{\rm circ}
 &=\frac{2\times4N_{\rm Golay}(9t_{\rm CZ}+t_M)}{p_s}
 \nonumber\\
 &\quad+N_{\rm Golay}(3\times2t_{\rm CZ}+2t_M)
 \nonumber\\
 &\simeq0.8155\ \mathrm{qubit}\mathbin{\cdot}\mathrm{s}.
 \label{eq:mb-cz-active-volume}
\end{align}
In each expression, the first term is the expected factory cost and the
second is the logical gate-teleportation cost; the first term for CZ prepares
two verified ancillas.  The factor $2t_{\rm CZ}$ counts two code blocks that
are simultaneously active during a transversal CZ layer, rather than two
sequential layers.  The MB-FTQC LERs in the tables are obtained by fitting
the numerical simulation results at multiple physical error rates and
evaluating the fits at $p=7.3\times10^{-4}$.

Preparing resource states offline reduces the MB-FTQC online critical path
to one CZ layer, one measurement layer, and a fixed-size parity lookup:
\begin{equation}
 T_{\rm MB,H}^{\rm circ}=T_{\rm MB,CZ}^{\rm circ}
 =t_{\rm CZ}+t_M=1.2\,\mathrm{ms}.
\end{equation}
A scalar benchmark on an Apple M2 processor gave a median LUT lookup time of
only $69.429\,\mathrm{ns}$.  Even our conservative $1\,\mu\mathrm{s}$ decoder
allowance increases the total latency only from $1.200$ to
$1.201\,\mathrm{ms}$ and is therefore negligible on the present timescale.
The LOBT operations on the two CZ input blocks run in parallel;
they increase the active volume but not the critical-path depth.
The decoder processes one error sector
and returns only the corrected logical parity through a
$2^{11}=2048$-entry LUT.  The lookup is fixed-size and deterministic.

Transversal STAR performs one SE round per logical gate and uses LogMat
correlated decoding.  A distance-$d$ rotated patch contains $ N_{\rm patch}=2d^2-1=97 $
physical qubits at $d=7$, where one patch comprises 49 data and 48 syndrome
qubits.  Its circuit costs are
\begin{align}
 T_{\rm 1SE,H}^{\rm circ}&=\tau_{\rm SE},
 \nonumber\\
 V_{\rm 1SE,H}^{\rm circ}
 &=N_{\rm patch}\tau_{\rm SE}
 =0.1746\ \mathrm{qubit}\mathbin{\cdot}\mathrm{s},
 \nonumber\\
 T_{\rm 1SE,CNOT}^{\rm circ}&=t_{\rm CZ}+\tau_{\rm SE},
 \nonumber\\
 V_{\rm 1SE,CNOT}^{\rm circ}
 &=2N_{\rm patch}(t_{\rm CZ}+\tau_{\rm SE})
 \nonumber\\
 &=0.3880\ \mathrm{qubit}\mathbin{\cdot}\mathrm{s}.
 \label{eq:transversal-star-cnot-active-volume}
\end{align}
The transversal physical-$H$ layer and patch reorientation are absorbed into
the adopted clock-cycle approximation.  The CNOT entangling layer is included
explicitly.  For LogMat, we adopt representative mean decoding times of about
$15\,\mu\mathrm{s}$ and $30\,\mu\mathrm{s}$ for the seven-gate $H$ and CNOT
circuits, respectively, in Fig.~10 of
Ref.~\cite{ismail2025transversalstararchitecturemegaquopscale}.  Dividing by
seven gives the per-gate decoding times used above.  For the latency
estimate, we use these per-gate averages to approximate the decoding time
for a single gate.  Adding these times and the associated
data-qubit holding volumes gives the totals in the tables.

For this comparison, we define the gate LER as the total probability of
nonidentity errors in the logical Pauli channel.  This quantity is not reported for
LogMat, so we estimate it using the MWPF$_{500}$ results in Table~III of the
same reference:
summing the nonidentity Pauli error probabilities gives $4\times10^{-7}$ for
$H$ and $1.8\times10^{-6}$ for the two-qubit gate~\footnote{The two-qubit entry
is labeled CZ in Table~III, whereas the main text describes CNOT extraction.
We use this entry as a reference for our CNOT estimate.}.
In Fig.~10, the LogMat circuit LERs are about 10 and 7 times the corresponding
MWPF$_{500}$ values for $H$ and CNOT, respectively.  Applying these ratios
gives estimated LogMat LERs of $4\times10^{-6}$ and $1.3\times10^{-5}$.
This assumes that the relative change in circuit LER between decoders also
approximately applies to the total Pauli error probability of each gate.

The repetitive-SE baseline uses $d=7$ SE rounds per gate with MWPM.
Replacing the single SE round by $d$ rounds, as in the $d$-round syndrome histories of
Ref.~\cite{Fowler2012}, gives
\begin{align}
 T_{\rm rep,H}^{\rm circ}&=d\tau_{\rm SE},
 \nonumber\\
 V_{\rm rep,H}^{\rm circ}
 &=N_{\rm patch}d\tau_{\rm SE}
 =1.2222\ \mathrm{qubit}\mathbin{\cdot}\mathrm{s},
 \nonumber\\
 T_{\rm rep,CNOT}^{\rm circ}&=t_{\rm CZ}+d\tau_{\rm SE},
 \nonumber\\
 V_{\rm rep,CNOT}^{\rm circ}
 &=2N_{\rm patch}(t_{\rm CZ}+d\tau_{\rm SE})
 \nonumber\\
 &=2.4832\ \mathrm{qubit}\mathbin{\cdot}\mathrm{s}.
\end{align}
For MWPM, we use $2\,\mu\mathrm{s}$ based on the $d=7$, $p=10^{-3}$
surface-code memory benchmark in Fig.~10 of the Sparse Blossom study
\cite{higgott2025sparseblossom}: the approximately
$0.2$--$0.3\,\mu\mathrm{s}$ per SE round corresponds to about
$2\,\mu\mathrm{s}$ for seven rounds.  For CNOT, the two patches are decoded
in parallel on two CPU cores, so the wall-clock decoding time remains
$2\,\mu\mathrm{s}$.

To obtain reference LERs, we use the threshold comparison in
Ref.~\cite{cain2025fastcorrelateddecodingtransversal}, which reports
$p_{\rm th}^{\rm corr}=0.718\%$ for a depth-14 transversal Clifford circuit
on 10 logical qubits with one SE round per layer, and
$p_{\rm th}^{\rm mem}=0.808\%$ for a single-logical-qubit memory with $d$ SE
rounds, both decoded with MWPM.  Applying the approximate scaling
$p_L\propto(p/p_{\rm th})^{\lfloor(d+1)/2\rfloor}$ to both benchmarks at
$d=7$, and assuming equal prefactors, gives the heuristic conversion factor
\begin{equation}
 \Gamma_7=
 \left(\frac{p_{\rm th}^{\rm mem}}{p_{\rm th}^{\rm corr}}\right)^4
 =\left(\frac{0.808}{0.718}\right)^4\simeq1.60.
\end{equation}
Assuming that this factor also applies to the gate-level total Pauli error
probabilities under the noise models used here, we take
\begin{equation}
 p_{\mathrm{tot},g}^{\rm rep}
 \sim\frac{p_{\mathrm{tot},g}^{\rm LogMat}}{\Gamma_7},
 \qquad g\in\{H,\mathrm{CNOT}\},
\end{equation}
giving $2.5\times10^{-6}$ for $H$ and $8\times10^{-6}$ for CNOT.
The source benchmarks use different failure observables and normalizations
and do not directly compare the two gate implementations.
The MB-FTQC LERs are upper-bound estimates,
whereas the LogMat and repetitive-SE values are approximate estimates
inferred from published benchmarks.
The LER comparison in Tables~\ref{tab:clifford-h-spacetime-comparison}
and~\ref{tab:clifford-cz-spacetime-comparison} should therefore be regarded
as approximate. In particular, the repetitive-SE reference LERs
should not be interpreted as established lower bounds.

\section{Non-Clifford resource-state spacetime-cost comparison with fold-transversal cultivation}
\label{app:nonclifford-fold-transversal-comparison}

This appendix derives the expected spacetime volumes per accepted proxy
resource state reported in Table~\ref{tab:nonclifford-spacetime-comparison}
of Sec.~\ref{subsec:surface-code-resource-comparison}, comparing Golay-code
HO-ZLD at $r=2$ and $r=3$ with the $f=5$ fold-transversal surface-code
cultivation protocol.

To compare the two postselected factories consistently, we use the same
proxy state $\ket{S}_L$ and the expected-volume convention of
Ref.~\cite{sahay2026foldtransversalcultivation}.  Both sets of results use
$p=10^{-3}$, but the channel assignments are not identical: HO-ZLD uses the
standard circuit-level depolarizing model of this work without idling faults,
whereas the fold-transversal data use the published uniform depolarizing model
SD6.  The logical accuracy is compared for the common proxy
$\ket{S}_L=S_L\ket{+}_L$; hence no factor-of-two conversion from the proxy-state
error to a $T$-state error is applied.  Let $C_i$ be the circuit segment after
the first $i$ post-selection stages, $V_i$ its active-qubit--gate-step volume,
$f_i$ the fraction of shots reaching that segment, with $f_0=1$, and $f_M$
the final acceptance probability.  The expected volume per accepted proxy
state is
\begin{equation}
 V_{\rm success}=\frac{1}{f_M}\sum_i f_iV_i.
 \label{eq:postselected-resource-volume}
\end{equation}
This weighting credits early termination when a rejected shot does not enter
later segments.  Following the fold-transversal convention, each elementary
gate, initialization, measurement, or idle layer counts as one step, with
explicit basis-changing gates counted separately.

For the fold-transversal reference, we stop the $f=5$ protocol immediately
after its final gap-based post-selection at $d=11$.  The subsequent unitary
growth to $d=15$ does not introduce another post-selection stage and is
omitted, thereby giving the fold-transversal baseline a smaller resource
estimate.  At this stopping point, changing the complementary-gap threshold
changes only the final acceptance probability; the weighted work preceding
that decision remains
$\sum_i f_iV_i=4{,}301.29$ qubit$\cdot$step, as obtained from the $f=5$
segment volumes in Table A5 and the shot-survival data of
Ref.~\cite{sahay2026foldtransversalcultivation}.  Interpolation of its
Fig.~2(a) gives expected attempt counts of $4.33$ and $4.94$ at the matched
LERs $1.27\times10^{-7}$ and $1.45\times10^{-8}$, respectively, corresponding
to $f_M\simeq0.2310$ and $0.2024$.  Thus, in units of
qubit$\cdot$step,
\begin{align}
 V_{\rm fold}(1.27\times10^{-7})
 &\simeq4{,}301.29\times4.33
 \simeq18{,}625,
 \nonumber\\
 V_{\rm fold}(1.45\times10^{-8})
 &\simeq4{,}301.29\times4.94
 \simeq21{,}248.
\end{align}

Applying the same segment-weighted accounting to HO-ZLD includes the
probabilistic preparation of every verified Golay $\ket{0}_L$ ancilla.  At
$p=10^{-3}$, its preparation succeeds with probability $p_s=0.6068$, so the
corresponding expected factory contribution is weighted by $1/p_s$.  The
Monte Carlo data underlying Figs.~\ref{fig:msd-error-rate}
and~\ref{fig:msd-success-rate} give
\begin{equation}
\begin{array}{c|cc}
 r & f_M & \sum_i f_iV_i\ (\text{qubit}\mathbin{\cdot}\text{step}) \\
 \hline
 2 & 0.612428 & 6{,}414.7 \\
 3 & 0.491997 & 8{,}559.2
\end{array}
\end{equation}
and therefore, again in units of qubit$\cdot$step,
\begin{align}
 V_{\rm HO\text{-}ZLD}^{(r=2)}
 &=\frac{6{,}414.7}{0.612428}
 \simeq10{,}474,
 \nonumber\\
 V_{\rm HO\text{-}ZLD}^{(r=3)}
 &=\frac{8{,}559.2}{0.491997}
 \simeq17{,}397.
\end{align}
These values differ from $C_{\rm avg}$ in
Table~\ref{tab:postselection-overhead}, which counts physical operations and
charges a rejected attempt for the full circuit; the present comparison uses
qubit$\cdot$step and credits segment-level early termination to match the cost
definition used for the fold-transversal protocol.

This is an aligned architecture-level comparison rather than a common
end-to-end simulation: the nominal physical-error scale, high-connectivity
assumption, proxy-state LER, and volume definition are matched, while the
native circuits and detailed noise channels remain different. In particular,
idling faults are omitted on the HO-ZLD side;
Appendices~\ref{app:memory-errors} and~\ref{app:memory-error-sensitivity}
indicate that their contribution is expected to be small for the target
trapped-ion and neutral-atom platforms.

\end{document}